\documentclass[10pt, journal]{IEEEtran}

\usepackage{graphicx, bm, mathrsfs}
\usepackage{cite,array}
\usepackage{amssymb}
\usepackage{dsfont}
\usepackage{epstopdf}
\usepackage[cmex10]{amsmath}
\usepackage{enumerate}
\usepackage{multirow}
\usepackage{booktabs}
\usepackage{threeparttable}

\newtheorem{assumption}{Assumption}
\newtheorem{theorem}{Theorem}
\newtheorem{corollary}{Corollary}

\newtheorem{proposition}{Proposition}
\newtheorem{definition}{Definition}
\newtheorem{remark}{Remark}

\DeclareMathOperator{\p}{p}
\DeclareMathOperator{\rf}{r}
\DeclareMathOperator{\ct}{c}
\DeclareMathOperator{\f}{f}
\DeclareMathOperator{\df}{d}
\DeclareMathOperator{\dom}{dom}
\DeclareMathOperator*{\esssup}{ess.\,sup}

\DeclareMathOperator{\mati}{mati}
\DeclareMathOperator{\mad}{mad}
\DeclareMathOperator{\diag}{diag}

\DeclareMathOperator{\tran}{^{\mkern-1.5mu\mathsf{T}}}

\begin{document}

\title{\LARGE \bf Tracking Control of Nonlinear Networked and Quantized Control Systems with Communication Delays
\thanks{This work was supported by National Natural Science Foundation of China, under Grant 61773357.}
}

\author{Wei~Ren, and Junlin~Xiong, \IEEEmembership{Member, IEEE}
\thanks{W. Ren is with Division of Decision and Control Systems, EECS, KTH Royal Institute of Technology, SE-10044, Stockholm, Sweden. J. Xiong is with Department of Automation, University of Science and Technology of China, Hefei, 230026, China.
Email: \texttt{\small weire@kth.se}, \texttt{\small junlin.xiong@gmail.com}.}
}

\maketitle

\begin{abstract}
This paper studies the tracking control problem of nonlinear networked and quantized control systems (NQCSs) with communication delays. The desired trajectory is generated by a reference system. The communication network is to guarantee the information transmission among the plant, the reference system and the controller. The communication network also brings about some undesired issues like time-varying transmission intervals, time-varying transmission delays, packet dropouts, scheduling and quantization effects, which lead to non-vanishing network-induced errors and affect the tracking performances. As a result, we develop a general hybrid system model for NQCSs with all aforementioned issues. Based on the Lyapunov approach, sufficient conditions are established to guarantee the stability of the tracking error with respect to the non-vanishing network-induced errors. The obtained conditions lead to a tradeoff between the maximally allowable transmission interval and the maximally allowable delay. Furthermore, the existence of Lyapunov functions satisfying the obtained conditions is studied. For specific time-scheduling protocols (e.g., Round-Robin protocol and Try-Once-Discard protocol) and quantizers (e.g., zoom quantizer and box quantizer), Lyapunov functions are constructed explicitly. Finally, a numerical example is presented to demonstrate the developed theory.
\end{abstract}


\section{Introduction}
\label{sec-introduction}

Because of fast development and widespread application of digital network technologies, networked control systems (NCSs) have attracted attention in the control community over the past decades; see \cite{Baillieul2007control, Lunze2014control, Gupta2010networked, Bauer2013networked, Antsaklis2007special}. The presence of the network offers considerable advantages over the traditional feedback control systems in terms of simplicity and flexibility in installation and maintenance, low cost and convenient resource sharing. On the other hand, the introduction of the limited-capacity network also induces many issues. The network-induced issues can be grouped into five types as follows \cite{Karafyllis2012nonlinear, Donkers2011stability, Nesic2009unified, Heemels2010networked}: time-varying transmission intervals; time-varying transmission delays; quantization errors; packet dropouts (caused by the unreliability of the network); and communication constraints (caused by the sharing of the network by multiple nodes and the fact that only one node is allowed to transmit its packet per transmission). Therefore, system modelling, stability analysis and controller design are fundamental problems for NCSs. Based on different modeling approaches and analysis methods, many results have been obtained in the literature. For instance, both system modelling and stability analysis have been studied in \cite{Carnevale2007lyapunov, Heemels2010networked, Nesic2004input, Nesic2009unified, Antunes2013stability}, and stabilizing controllers have been developed in \cite{Karafyllis2012nonlinear, Cloosterman2010controller, Gao2008network, Hirche2009distributed}.

However, as another fundamental problem in control theory, tracking control has seldom been studied for NCSs; see \cite{Van2010tracking, Gao2008network, Zhang2013step, Postoyan2014tracking}. The main objective of tracking control is to design appropriate controller such that the considered system can track a given reference trajectory as close as possible; see \cite{Biemond2013tracking, Lian2013robust, Grimble2005nonlinear}. In the tracking control, the controller consists of two parts \cite{Van2007tracking, Van2010tracking}: the feedforward part to induce the reference trajectory in the whole system, and the feedback part to ensure the stabilization of the considered system and the convergence to the reference trajectory. Compared with stability analysis, the tracking control problem is well recognized to be more general and more difficult \cite{Garcia2011model, Gao2008network}. In addition, due to the presence of the network, the aforementioned network-induced issues have great impacts on the tracking performance. For instance, both quantization and time-varying transmission delays result in the feedforward error; the communication constraints and limited capacity of the network deteriorate the tracking performance. Therefore, only approximate tracking can be achieved \cite{Van2010tracking}. For instance, the approximate tracking control problem has been studied in \cite{Van2010tracking} for sampled-data systems, and in \cite{Postoyan2014tracking} for NCSs with time-varying transmission intervals and delays via the emulation-like approach as in \cite{Heemels2010networked, Nesic2004input}. However, observe from all the previous works that only parts of the aforementioned issues are studied, which motivates us to study this topic further.

In this paper, we study the tracking control problem for nonlinear networked and quantized control systems (NQCSs) with communication delays, which are nonlinear NCSs with all the aforementioned issues.
To this end, a unified hybrid model in the formalism of \cite{Cai2009characterizations, Goebel2006solutions} is developed for the tracking control of NQCSs with communication delays based on the emulation-like approach as in \cite{Carnevale2007lyapunov, Nesic2009unified, Heemels2010networked}, which is our first contribution. We aim at proposing a high fidelity model that is amenable to controller design and tracking performance analysis. To achieve this, all aforementioned network-induced issues are studied, including time-varying transmission intervals; time-varying transmission delays and quantization errors; packet dropouts and communication constraints. In addition, a general quantizer is proposed to recover most types of the quantizers in previous works \cite{Liberzon2003stabilization, Brockett2000quantized, Liberzon2003hybrid, Nesic2009unified}. As a result, the proposed hybrid model extends those in previous works \cite{Gao2008network, Postoyan2014tracking, Van2010tracking, Nesic2009unified, Heemels2010networked} on stability analysis and tracking control of NCSs.

Our second contribution is to establish sufficient conditions to guarantee the convergence of the tracking error with respect to network-induced errors using the Lyapunov-based approach. To this end, some reasonable assumptions are provided, which are different from those in \cite{Heemels2010networked} for stability analysis of NCSs. With these assumptions, We derive the tradeoff between the maximally allowable transmission interval (MATI) and the maximally allowable delay (MAD) to guarantee that the tracking error converges to the origin up to some errors due to the aforementioned network-induced errors. These network-induced errors not only leads to the main difference from the scenario of stabilizing an equilibrium point, but also results in additional technical difficulties in tracking performance analysis. Note that the obtain results are also available for the scenario of stabilizing an equilibrium point. In addition, the tradeoff depends on the applied communication protocol, and thus allows for the comparison of different protocols.

Since the assumptions we adopt are different from those for stability analysis, it is necessary to verify the existence of Lyapunov functions satisfying these assumptions, which is the third contribution of this paper. The construction of Lyapunov functions is presented explicitly based on the quantization-free and delay-free case in \cite{Postoyan2014tracking}, the quantization-free case \cite{Heemels2010networked} and the delay-free cases in \cite{Nesic2009unified, Carnevale2007lyapunov}. In addition, for different time-scheduling protocols and quantizers, specific Lyapunov functions are established. In the construction of Lyapunov functions, we also show how to reduce the effects of the network-induced errors on the tracking performance through the implementation of the controller and the design of the time-scheduling protocol.

A preliminary version of this work has been presented in the conference paper \cite{Ren2018tracking} where the zoom quantizer is considered and the reference trajectory is required to be convergent. The current paper extends the approach to consider general NQCSs with a more general quantizer and no constraints on the reference trajectory. In addition, Lyapunov functions is constructed explicitly in this paper. Therefore, the result of \cite{Ren2018tracking} is recovered as a particular case.

This paper is organized as follows. Preliminaries are presented in Section \ref{sec-preliminaries}. In Section \ref{sec-problemformation}, the tracking problem is formulated and a unified system model is developed. The Lyapunov-based conditions are obtained in Section \ref{sec-mainresults} to guarantee the convergence of the tracking error and the tradeoff between the MATI and the MAD. The existence of Lyapunov functions is studied in Section \ref{sec-lyapunovfunction}. In Section \ref{sec-illustration}, the developed results are illustrated by a numerical example. Conclusions and further researches are stated in Section \ref{sec-conclusion}.

\section{Preliminaries}
\label{sec-preliminaries}

Basic definitions and notation are presented in this section. $\mathbb{R}:=(-\infty, +\infty)$; $\mathbb{R}_{\geq0}:=[0, +\infty)$; $\mathbb{R}_{>0}:=(0, +\infty)$; $\mathbb{N}:=\{0, 1, 2, \ldots\}$; $\mathbb{N}_{>0}:=\{1, 2, \ldots\}$. Given two sets $\mathcal{A}$ and $\mathcal{B}$, $\mathcal{B}\backslash\mathcal{A}:=\{x | x\in\mathcal{B}, x\notin\mathcal{A}\}$. Given a constant $a\in\mathbb{R}$ and a set $\mathcal{A}$, $a\mathcal{A}:=\{ax| x\in\mathcal{A}\}$. A set $\mathcal{A}\subseteq\mathbb{R}^{n}$ is symmetric if $-x\in\mathcal{A}$ for all $x\in\mathcal{A}$. $|\cdot|$ stands for Euclidean norm; $\|\cdot\|_{\mathfrak{J}}$ denotes the supremum norm of a function on an interval $\mathfrak{J}$ and $\|\cdot\|$ denotes the supremum norm in the case of $\mathfrak{J}=[t_{0}, \infty)$, where $t_{0}\in\mathbb{R}_{\geq0}$ is the given initial time. For the vectors $x, y\in\mathbb{R}^{n}$, $(x, y):=(x\tran, y\tran)\tran$ for simplicity of notation and $\langle x, y\rangle$ denotes the usual inner product. $I_{n}$ represents the identity matrix of dimension $n$, and $\diag\{A, B\}$ denotes the block diagonal matrix made of the square matrices $A$ and $B$. The symbols $\wedge$ and $\vee$ denote separately `and' and `or' in logic. $\bm{B}(a, b)$ denotes the hypercubic box centered at $a\in\mathbb{R}^{n}$ with edges of length $2b$. $f(t^{+}):=\limsup_{s\rightarrow0^{+}}f(t+s)$ for a given function $f: \mathbb{R}_{\geq t_{0}}\rightarrow\mathbb{R}^{n}$. A function $\alpha: \mathbb{R}_{\geq0}\rightarrow\mathbb{R}_{\geq0}$ is of class $\mathcal{K}$ if it is continuous, $\alpha(0)=0$, and strictly increasing; it is of class $\mathcal{K}_{\infty}$ if it is of class $\mathcal{K}$ and unbounded. A function $\beta: \mathbb{R}_{\geq0}\times\mathbb{R}_{\geq0}\rightarrow\mathbb{R}_{\geq0}$ is of class $\mathcal{KL}$ if $\beta(s, t)$ is of class $\mathcal{K}$ for each fixed $t\geq0$ and $\beta(s, t)$ decreases to zero as $t\rightarrow0$ for each fixed $s\geq0$. A function $\beta:\mathbb{R}_{\geq0}\times\mathbb{R}_{\geq0}\times\mathbb{R}_{\geq0}\rightarrow\mathbb{R}_{\geq0}$ is of class $\mathcal{KLL}$ if $\beta(r, s, t)$ is of class $\mathcal{KL}$ for each fixed $s\geq0$ and of class $\mathcal{KL}$ for each fixed $t\geq0$.

The basic concepts of hybrid systems are introduced as follows; see \cite{Cai2009characterizations} for the details. Consider hybrid systems of the form:
\begin{equation}
\label{eqn-1}
\begin{cases}
\dot{x}=F(x, w), \quad (x, w)\in C; \\
x^{+}=G(x, w), \quad (x, w)\in D,
\end{cases}
\end{equation}
where $x\in\mathbb{R}^{n}$ is the system state, $w\in\mathbb{R}^{m}$ is the external input, $F: C\rightarrow\mathbb{R}^{n}$ is the flow map, $G: D\rightarrow\mathbb{R}^{m}$ is the jump map, $C$ is the flow set and $D$ is the jump set. For the hybrid system \eqref{eqn-1}, the following basic assumptions are given \cite{Cai2009characterizations}: the sets $C, D\subset\mathbb{R}^{n}\times\mathbb{R}^{m}$ are closed; $F$ is continuous on $C$; and $G$ is continuous on $D$.

A subset $E\subset\mathbb{R}_{\geq0}\times\mathbb{N}$ is a \emph{compact hybrid time domain} if $E=\bigcup_{0\leq j\leq J}([t_{j}, t_{j+1}], j)$ for some finite sequence of times $0=t_{0}\leq t_{1}\leq\ldots\leq t_{J+1}$. $E$ is a \emph{hybrid time domain} if for all $(T, J)\in E$, $E\bigcap([0, T]\times\{0, \ldots, J\})$ is a compact hybrid time domain. For all $(t_{1}, j_{1}), (t_{2}, j_{2})\in\mathbb{R}_{\geq0}\times\mathbb{N}$, denote $(t_{1}, j_{1})\preceq(t_{2}, j_{2})$ (or $(t_{1}, j_{1})\prec(t_{2}, j_{2})$) if $t_{1}+j_{1}\leq t_{2}+j_{2}$ (or $t_{1}+j_{1}<t_{2}+j_{2}$). A function $w: \dom w\rightarrow\mathbb{R}^{m}$ is a \emph{hybrid input} if $w(\cdot, j)$ is Lebesgue measurable and locally essentially bounded for each $j$. A function $x: \dom x\rightarrow\mathbb{R}^{n}$ is a \emph{hybrid arc} if $x(\cdot, j)$ is locally absolutely continuous for each $j$. The hybrid arc $x: \dom x\rightarrow\mathbb{R}^{n}$ and the hybrid input $w: \dom w\rightarrow\mathbb{R}^{m}$ are a \emph{solution pair} to \eqref{eqn-1} if: \romannumeral1) $\dom x=\dom w$ and $(x(t_{0},j_{0}), w(t_{0},j_{0}))\in C\cup D$; \romannumeral2) for all $j\in\mathbb{N}$ and almost all $t$ such that $(t, j)\in\dom x$, $(x(t, j), w(t, j))\in C$ and $\dot{x}(t, j)=F(x(t, j), w(t, j))$; \romannumeral3) for all $(t, j)\in\dom x$ such that $(t, j+1)\in\dom x$, $(x(t, j), w(t, j))\in D$ and $x(t, j+1)=G(x(t, j), w(t, j))$. A solution pair $(x, u)$ to (1) is \emph{maximal} if it cannot be extended, and it is \emph{complete} if $\dom x$ is unbounded. Let $w$ be a hybrid input with $(0, 0)$ as initial hybrid time, we define $\|w\|_{(t, j)}:=\max\left\{\esssup\limits_{(t', j')\in\dom w\setminus\Gamma(w), (0, 0)\preceq(t', j')\preceq(t, j)}|w(t', j')|,\right.$ $\left.\sup\limits_{(t, j)\in\Gamma(w), (0, 0)\preceq(t', j')\preceq(t, j)}\sup|w(t', j')|\right\}$ where $\Gamma(w)$ denotes the set of all $(t, j)\in\dom w$ such that $(t, j+1)\in\dom w$. Denote by $\mathfrak{S}_{w}(x_{0})$ the set of all maximal solution pairs $(x, w)$ to the system \eqref{eqn-1} with $x_{0}\in C\cup D$ and finite $\|w\|:=\sup_{(t, j)\in\dom w, t+j\rightarrow\infty}\|w\|_{(t, j)}$.

\begin{definition}[\cite{Cai2009characterizations}]
The hybrid system \eqref{eqn-1} is \emph{input-to-state stable (ISS) } from $w$ to $x$, if there exist $\beta\in\mathcal{KLL}, \gamma\in\mathcal{K}_{\infty}$ such that for all $(t, j)\in\dom x$ and all $(x, w)\in\mathfrak{S}_{w}(x(0, 0))$,
\begin{equation*}
|x(t, j)|\leq\beta(|x(0, 0)|,t, j)+\gamma(\|w\|_{(t, j)}).
\end{equation*}
In addition, $\beta(v, t, j)=Kve^{-(t+j)}$, where $K>0$, then the system \eqref{eqn-1} is \emph{exponentially input-to-state stable (EISS) from $w$ to $x$}.
\end{definition}

\section{Problem Formulation}
\label{sec-problemformation}

In this section, the tracking control problem for NQCSs with communication delays is formulated using the emulation approach as proposed in \cite{Nesic2004input, Carnevale2007lyapunov}.

\subsection{Tracking Problem of NQCSs}
\label{subsec-trackingproblem}

Consider the nonlinear system of the form
\begin{align}
\label{eqn-2}
\dot{x}_{\p}=f_{\p}(x_{\p}, u),\quad y_{\p}=g_{\p}(x_{\p}),
\end{align}
where $x_{\p}\in\mathbb{R}^{n_{\p}}$ is the system state, $u\in\mathbb{R}^{n_{u}}$ is the control input, and $y_{\p}\in\mathbb{R}^{n_{y_{\p}}}$ is the system output. The reference system tracked by the system \eqref{eqn-2} is of the form:
\begin{align}
\label{eqn-3}
\dot{x}_{\rf}=f_{\p}(x_{\rf}, u_{\f}),\quad y_{\rf}=g_{\p}(x_{\rf}),
\end{align}
where $x_{\rf}\in\mathbb{R}^{n_{\rf}}$ is the reference state ($n_{\rf}=n_{\p}$), $u_{\f}\in\mathbb{R}^{n_{u}}$ is the feedforward control input, and $y_{\rf}\in\mathbb{R}^{n_{y_{\rf}}}$ is the reference output ($n_{y_{\rf}}=n_{y_{\p}}=n_{y}$). Assume that the reference system \eqref{eqn-3} has a unique solution for any initial condition and any input.

To track the reference system, the controller, which is designed for \eqref{eqn-2} in the absence of the network, is given by
\begin{equation}
\label{eqn-4}
u=u_{\ct}+u_{\f},
\end{equation}
where $u_{\f}\in\mathbb{R}^{n_{u}}$ is the feedforward item, and $u_{\ct}\in\mathbb{R}^{n_{u}}$ is the feedback item. The feedback item $u_{\ct}$ is from the nonlinear feedback controller given below:
\begin{align}
\label{eqn-5}
\dot{x}_{\ct}=f_{\ct}(x_{\ct}, y_{\p}-y_{\rf}),\quad u_{\ct}=g_{\ct}(x_{\ct}),
\end{align}
where $x_{\ct}\in\mathbb{R}^{n_{\ct}}$ is the feedback controller state, $u_{\ct}\in\mathbb{R}^{n_{u}}$ is the feedback controller output. Observe from \eqref{eqn-5} that the feedback controller depends on the difference between the outputs $y_{\p}$ and $y_{\rf}$, which is different from the cases studied in previous works \cite{Carnevale2007lyapunov, Nesic2009unified, Nesic2004input, Heemels2010networked, Postoyan2014tracking} where the feedback controller depends on the outputs $y_{\p}$ and $y_{\rf}$. The feedback controller of the form \eqref{eqn-5} can be found in the literature \cite{Lian2013robust, Van2010tracking}. In addition, denote $y_{\df}:=y_{\p}-y_{\rf}\in\mathbb{R}^{n_{y}}$ for the sake of convenience.

\begin{figure}[!t]
\begin{center}
\begin{picture}(100,125)
\put(-50,-12){\resizebox{70mm}{45mm}{\includegraphics[width=2.5in]{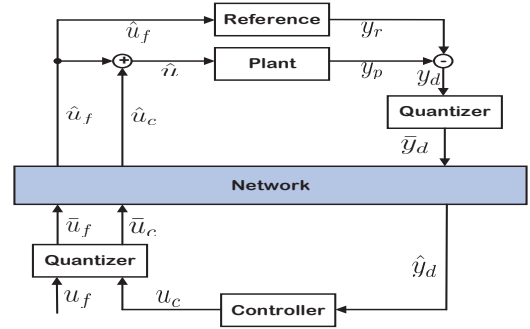}}}
\end{picture}
\end{center}
\caption{General framework of tracking control of networked and quantized control systems (NQCSs).}
\label{fig-1}
\end{figure}

Assume that $f_{\p}$ and $f_{\ct}$ are continuous; $g_{\p}$ and $g_{\ct}$ are continuously differentiable. The objective of this paper is to implement the designed controller over the network, as illustrated in Fig. \ref{fig-1}, and to demonstrate that under reasonable assumptions, the assumed tracking performance of the system \eqref{eqn-2}-\eqref{eqn-5} will be preserved for the NQCS.

\subsection{Information Transmission over Quantizer and Network}
\label{subsec-informationtransmission}

At the transmission times $t_{s_{i}}$, $i\in\mathbb{N}$, (part of) the outputs of the plant, the reference system and the controller are sampled, quantized and transmitted through the communication network. The communication network is used to guarantee the information transmission among the sensors, the controller and the actuators. Based on the band-limited network and the spatial location of the sensors and actuators \cite{Tabbara2007stability}, we group the sensors and actuators into $l\in\mathbb{N}_{>0}$ nodes connecting the network. Correspondingly, the transmitted information is partitioned into $l$ parts. At each $t_{s_{i}}$, one and only one node is allowed to access to the network, which is determined by time-scheduling protocols; see also \cite{Donkers2011stability, Nesic2009unified, Nesic2004input} and Subsection \ref{subsec-protocol}. The transmission times are strictly increasing, and the transmission intervals are defined as $h_{i}:=t_{s_{i+1}}-t_{s_{i}}$, $i\in\mathbb{N}$. Due to computation and data coding, the information can not be transmitted instantaneously; see \cite{Liu2015dynamic, Wang2015emulation}. As a result, there exist transmission delays $\tau_{i}\geq0$, $i\in\mathbb{N}$, such that the controller and actuators receive the transmitted information at the arrival times $r_{i}=t_{s_{i}}+\tau_{i}$, $i\in\mathbb{N}$. For both $h_{i}$ and $\tau_{i}$, $i\in\mathbb{N}$, the following assumption is adopted; see also \cite[Assumption \uppercase\expandafter{\romannumeral2}.1]{Heemels2010networked}.

\begin{assumption}
\label{asn-1}
There exist constants $h_{\mati}\geq h_{\mad}\geq0$ and $\varepsilon\in(0, h_{\mati})$ such that $\varepsilon\leq h_{i}\leq h_{\mati}$ and $0\leq\tau_{i}\leq\min\{h_{\mad}, h_{i}\}$ for all $i\in\mathbb{N}$.
\end{assumption}

In Assumption \ref{asn-1}, $h_{\mati}$ and $h_{\mad}$ are called the \textit{maximally allowable transfer interval (MATI)} and the \textit{maximally allowable delay (MAD)}, respectively. $\varepsilon>0$ implies that there are no Zeno solutions for the closed-loop system; see \cite[Assumption 1]{Nesic2009unified}, \cite{Wang2015emulation}. Assumption \ref{asn-1} guarantees that each transmitted information arrives before the next sampling, which is called the small delay case; see \cite{Van2013stability, Heemels2010networked}.

\begin{remark}
\label{rmk-1}
In Assumption \ref{asn-1}, $\varepsilon>0$ always exists in real networks and is called the minimum inter-transmission interval; see \cite{Wang2015emulation}. If packet dropouts are considered, then $h_{\mati}$ is adjusted as $\bar{h}_{\mati}=h_{\mati}/(\aleph+1)$, where $\aleph\in\mathbb{N}$ is the maximal number of successive packet dropouts; see also \cite{Heemels2010networked}.
\hfill $\square$
\end{remark}

As aforementioned, to match the limited transmission capacity of the network, $y_{\df}, u_{\ct}$ and $u_{\f}$ are quantized before they are transmitted via the network. Each node has a quantizer, which is a piecewise continuous function $q_{j}: \mathbb{R}_{>0}\times\mathbb{R}^{n_{j}}\rightarrow\mathcal{Q}_{j}\subset\mathbb{R}^{n_{j}}$, where $\mathcal{Q}_{j}$ is a finite or countable set and $j\in\{1, \ldots, l\}$. The quantizer has a quantization parameter $\mu_{j}>0$ and satisfies the following assumption, which is a generalization of those in \cite{Liberzon2007input, Liberzon2003stabilization, Liberzon2003hybrid}.

\begin{assumption}
\label{asn-2}
For each $j\in\{1, \ldots, l\}$, there exist nonempty symmetric sets $\mathds{C}_{0j}, \mathds{C}_{j}, \mathds{D}_{j}\subset\mathbb{R}^{n_{j}}$ and a constant $d_{j}\in[0, 1)$ such that $\mathds{C}_{0j}\subseteq\mathds{C}_{j}$, the origin is contained in $\mathds{C}_{0j}, \mathds{C}_{j}, \mathds{D}_{j}$ and for all $z_{j}\in\mathbb{R}^{n_{j}}$,
\begin{align}
\label{eqn-6}
z_{j}\in\mathds{C}_{j}&\Rightarrow q_{j}(\mu_{j}, z_{j})-z_{j}\in\mathds{D}_{j}, \\
\label{eqn-7}
z_{j}\notin\mathds{C}_{j}&\Rightarrow q_{j}(\mu_{j}, z_{j})\notin(1-d_{j})\mathds{C}_{j}, \\
\label{eqn-8}
z_{j}\in\mathds{C}_{0j}&\Rightarrow q_{j}(\mu_{j}, z_{j})\equiv0.
\end{align}
\end{assumption}

In Assumption \ref{asn-2}, $\epsilon_{j}:=q_{j}(\mu_{j}, z_{j})-z_{j}$ is defined as the quantization error. For each $j\in\{1, \ldots, l\}$, $\mathds{C}_{j}$ is the union of all the quantization regions; $\mathds{D}_{j}$ is the set of all the (possible) quantization errors; $\mathds{C}_{0j}$ is the deadzone of the quantizer. Condition \eqref{eqn-6} gives a bounded region on the quantization error if the quantizer does not saturate. Condition \eqref{eqn-7} provides an approach to detect the possible saturation. Note that the constant $d_{j}$ is related to the sizes of the sets $\mathds{C}_{j}$ and $\mathds{D}_{j}$. If the signal is so small, then condition \eqref{eqn-8} implies that it is reasonable to quantize such signal as zero. Two special quantizers will be studied in Subsections \ref{subsec-zoomquantizer} and \ref{subsec-boxquantizer}. On the other hand, assume that the quantizer is applied synchronously on both sides of the network; otherwise, see \cite{Kameneva2009robustness} for more details.

\begin{remark}
\label{rmk-2}
The applied quantizer is a dynamic quantizer due to the quantization parameter. Because of the applied dynamic quantizer, the feedback control is of the form \eqref{eqn-5} such that $y_{\df}$ is driven to converge. If the reference trajectory is convergent, then the convergence point of the reference trajectory can be set as the origin in the applied quantizer. In this case, we can relax the feedback controller \eqref{eqn-5} as the following form:
\begin{align}
\label{eqn-9}
\dot{x}_{\ct}=f_{\ct}(x_{\ct}, y_{\p}, y_{\rf}),\quad u_{\ct}=g_{\ct}(x_{\ct}).
\end{align}
That is, the controller depends on the outputs $y_{\p}$ and $y_{\rf}$, which is similar to the cases in previous works \cite{Carnevale2007lyapunov, Nesic2009unified, Nesic2004input, Heemels2010networked, Postoyan2014tracking}. In this case, the system modelling and tracking performance analysis in this paper can be proceeded along the similar fashion with a slight modification; see also \cite{Ren2018tracking, Postoyan2014tracking}.
\hfill $\square$
\end{remark}

\begin{remark}
\label{rmk-3}
The quantizer satisfying Assumption \ref{asn-2} is an extension of zoom quantizer in \cite{Liberzon2003hybrid} (see also Subsection \ref{subsec-zoomquantizer}) and includes many types of quantizers in the existing works \cite{Brockett2000quantized, Fu2005sector, Liberzon2003stabilization, Van2013stability}. For instance, for the uniform quantizer in \cite{Brockett2000quantized, Liberzon2003hybrid}, $\mu_{j}$ is constant, $\mathds{C}_{j}=\{z_{j}\in\mathbb{R}^{n_{j}}||z_{j}|\leq M_{j}\}$, $\mathds{D}_{j}=\{\epsilon_{j}\in\mathbb{R}^{n_{j}}||\epsilon_{j}|\leq\Delta_{j}\}$ and $d_{j}=\Delta_{j}/M_{j}$, where $M_{j}>\Delta_{j}>0$. For the zoom quantizer \cite{Liu2015dynamic, Liberzon2003hybrid}, $\mu_{j}$ is time-varying, $\mathds{C}_{j}=\{z_{j}\in\mathbb{R}^{n_{j}}||z_{j}|\leq M_{j}\mu_{j}\}$, $\mathds{D}_{j}=\{\epsilon_{j}\in\mathbb{R}^{n_{j}}||\epsilon_{j}|\leq\Delta_{j}\mu_{j}\}$ and $d_{j}=\Delta_{j}/M_{j}$, where $M_{j}>\Delta_{j}>0$. For the box quantizer in \cite{Nesic2009unified, Liberzon2003stabilization}, $\mathds{C}_{j}=\{z_{j}\in\mathbb{R}^{n_{j}}|z_{j}\in\bm{B}(\hat{z}_{j}, \mu_{j})\}$, $\mathds{D}_{j}=\{\epsilon_{j}\in\mathbb{R}^{n_{j}}||\epsilon_{j}|\leq\sqrt{n_{j}}\mu_{j}/N_{j}\}$ and $d_{j}=0$, where $\hat{z}_{j}$ is the estimate of $z_{j}$ and $N_{j}$ is a given constant. Note that $\mathds{C}_{0j}=\{z_{j}\in\mathbb{R}^{n_{j}}||z_{j}|\leq\Delta_{0j}\}$ for all the preceding quantizers and $\Delta_{0j}\geq0$ is a small constant.
\hfill $\square$
\end{remark}

All the quantization parameters in $l$ nodes are combined as $\mu:=(\mu_{1}, \ldots, \mu_{l})\in\mathbb{R}^{l}_{>0}$, and then the overall quantizer is defined as
\begin{equation*}
q(\mu, z):=(q_{1}(\mu_{1}, z_{1}), \ldots, q_{l}(\mu_{l}, z_{l})).
\end{equation*}
The quantization parameter $\mu\in\mathbb{R}^{l}_{>0}$ evolves according to a hybrid dynamics, which will be given later. The quantized measurements are defined as $\bar{y}_{\df}:=q(\mu, y_{\df})$, $\bar{u}_{\ct}:=q(\mu, u_{\ct})$ and $\bar{u}_{\f}:=q(\mu, u_{\f})$. Correspondingly, the quantization errors are $\epsilon_{\df}:=\bar{y}_{\df}-y_{\df}$, $\epsilon_{\ct}:=\bar{u}_{\ct}-u_{\ct}$ and $\epsilon_{\f}:=\bar{u}_{\f}-u_{\f}$. Denote $\bm{\epsilon}:=(\epsilon_{\df}, \epsilon_{\ct}, \epsilon_{\f})\in\mathbb{R}^{n_{y}+2n_{u}}$.

The quantized measurements are transmitted via the network and received at arrival times $r_{i}\in\mathbb{R}_{\geq0}$, $i\in\mathbb{N}$. In the following, we use the variables $\hat{u}_{\ct}, \hat{u}_{\f}\in\mathbb{R}^{n_{u}}$ and $\hat{y}_{\df}\in\mathbb{R}^{n_{y}}$ respectively denote the networked versions of $u_{\ct}, u_{\f}$ and $y_{\df}$. Therefore, the plant \eqref{eqn-2} receives $\hat{u}:=\hat{u}_{\ct}+\hat{u}_{\f}$, which is the networked version of the control input $u$; the reference system \eqref{eqn-3} receives $\hat{u}_{\f}\in\mathbb{R}^{n_{u}}$; and the feedback controller \eqref{eqn-5} receives $\hat{y}_{\df}\in\mathbb{R}^{n_{y}}$. In addition, the errors induced by the quantizer and the network are defined as $e_{\df}:=\hat{y}_{\df}-y_{\df}$, $e_{\ct}:=\hat{u}_{\ct}-u_{\ct}$ and $e_{\f}:=\hat{u}_{\f}-u_{\f}$. After the reception of the quantized measurements at arrival times, the received measurements are updated with the latest quantized measurements as follows.
\begin{align*}
\hat{y}_{\df}(r^{+}_{i})&=\bar{y}_{\df}(t_{s_{i}})+\bm{h}_{\df}(i, e_{\df}(t_{s_{i}}), e_{\ct}(t_{s_{i}}), e_{\f}(t_{s_{i}})),\\
\hat{u}_{\ct}(r^{+}_{i})&=\bar{u}_{\ct}(t_{s_{i}})+\bm{h}_{\ct}(i, e_{\df}(t_{s_{i}}), e_{\ct}(t_{s_{i}}), e_{\f}(t_{s_{i}})),\\
\hat{u}_{\f}(r^{+}_{i})&=\bar{u}_{\f}(t_{s_{i}})+\bm{h}_{\f}(i, e_{\df}(t_{s_{i}}), e_{\ct}(t_{s_{i}}), e_{\f}(t_{s_{i}})),
\end{align*}
where $\bm{h}_{\df}$, $\bm{h}_{\ct}$ and $\bm{h}_{\f}$ are the update functions, and depend on the time-scheduling protocol that determines which node is granted to access to the network; see also Subsection \ref{subsec-protocol}.

In the arrival intervals, the received measurements are assumed to be operated in zero-order hold (ZOH) fashion, i.e., for all $t\in(r_{i}, r_{i+1})$, $i\in\mathbb{N}$,
\begin{equation}
\label{eqn-10}
\dot{\hat{y}}_{\df}=0, \quad \dot{\hat{u}}_{\ct}=0, \quad \dot{\hat{u}}_{\f}=0.
\end{equation}
Therefore, at the arrival times $r_{i}$, $i\in\mathbb{N}$, the error $e_{\df}$ is updated as follows.
\begin{align}
\label{eqn-11}
e_{\df}(r^{+}_{i})&=\hat{y}_{\df}(r^{+}_{i})-y_{\df}(r^{+}_{i}) \nonumber\\
&=\bar{y}_{\df}(t_{s_{i}})+\bm{h}_{\df}(i, \bm{\vartheta}(t_{s_{i}}))-y_{\df}(r_{i}) \nonumber\\
&=e_{\df}(r_{i})-e_{\df}(t_{s_{i}})+\epsilon_{\df}(t_{s_{i}})+\bm{h}_{\df}(i, \bm{\vartheta}(t_{s_{i}}))\nonumber\\
&=:e_{\df}(r_{i})-e_{\df}(t_{s_{i}})+h_{\df}(i, y_{\df}(t_{s_{i}}), \bm{\vartheta}(t_{s_{i}}), \mu(t_{s_{i}})),
\end{align}
where $h_{\df}(i, y_{\df}, \bm{\vartheta}, \mu):=\epsilon_{\df}+\bm{h}_{\df}(i, \bm{\vartheta})$, $\bm{\vartheta}:=(e_{\df}, e_{\ct}, e_{\f})\in\mathbb{R}^{n_{\bm{\vartheta}}}$ and $n_{\bm{\vartheta}}=n_{y}+n_{\ct}+n_{\f}$. In \eqref{eqn-11}, the third ``$=$" holds due to ZOH device and $\hat{y}_{\df}(t_{s_{i}})=\hat{y}_{\df}(r_{i})$. Similarly,
\begin{align*}
& e_{\ct}(r^{+}_{i})=e_{\ct}(r_{i})-e_{\ct}(t_{s_{i}})+h_{\ct}(i, x_{\ct}(t_{s_{i}}), \bm{\vartheta}(t_{s_{i}}), \mu(t_{s_{i}})),\\
& e_{\f}(r^{+}_{i})=e_{\f}(r_{i})-e_{\f}(t_{s_{i}})+h_{\f}(i, x_{\f}(t_{s_{i}}), \bm{\vartheta}(t_{s_{i}}), \mu(t_{s_{i}})),
\end{align*}
where $x_{\f}$ is an auxiliary variable to ensure that the update of $e_{\f}$ is of the form \eqref{eqn-11}. The variable $x_{\f}$ is related to $u_{\f}$ because $h_{\f}(i, x_{\f}, \bm{\vartheta}, \mu)=q(\mu, u_{\f})-u_{\f}+\bm{h}_{\f}(i, \bm{\vartheta})$.

\begin{remark}
\label{rmk-4}
In this paper, the ZOH technique is required in \eqref{eqn-10}. The reason lies in the existence of time delays. For the time-delay case (see also \cite{Van2010tracking, Heemels2009networked, Heemels2010networked}), the ZOH technique leads to the fact that $\hat{y}_{\df}(t_{i})\equiv\hat{y}_{\df}(r_{i})$, which is applied in \eqref{eqn-11}. For the delay-free case (see also \cite{Nesic2004input, Nesic2009unified, Postoyan2014tracking}), such a technique is not required, and the generation of $(\hat{y}_{\df}, \hat{u}_{\ct}, \hat{u}_{\f})$ in \eqref{eqn-10} can be more flexible.
\hfill $\square$
\end{remark}

Similar to the evolution of the received measurements, the quantized measurements are operated in ZOH fashion on $(r_{i}, r_{i+1})$ and updated at $r_{i}$, $i\in\mathbb{N}$, with the update of $\mu\in\mathbb{R}^{l}_{>0}$. The evolution of $\mu$ is given as follows.
\begin{align}
\label{eqn-12}
\dot{\mu}(t)&=\bar{g}_{\mu}(y_{\df}, x_{\rf}, x_{\ct}, \bm{\vartheta}, \mu), \quad t\in[t_{s_{i}}, t_{s_{i+1}}]\setminus\{r_{i}\},\\
\label{eqn-13}
\mu(r^{+}_{i})&=\mu(r_{i})-\mu(t_{s_{i}})+h_{\mu}(i, \mu(t_{s_{i}}), \bm{\epsilon}(t_{s_{i}})),
\end{align}
where $\bar{g}_{\mu}$ is an evolution function and $h_{\mu}$ is an update function that depends on the time-scheduling protocol.

\begin{remark}
\label{rmk-5}
The evolution of the quantization parameter $\mu$ is analogous to those in \cite{Tabbara2008input, Nesic2009unified, Heemels2009networked}. That is, $\mu$ is time-varying in continuous intervals and updated at discrete-time instants. If $\mu$ is time-invariant in $[t_{s_{i}}, t_{s_{i+1}})$ and only updated at the arrival times, then $\dot{\mu}\equiv0$ in $(r_{i}, r_{i+1})$ and $\mu(r^{+}_{i})=h_{\mu}(i, \mu(t_{s_{i}}), \bm{\epsilon}(t_{s_{i}}))$. This scenario is similar to the case in \cite{Heemels2009networked}. However, $\mu$ is updated at the arrival times in this paper instead of at the transmission times in all the previous works \cite{Tabbara2008input, Nesic2009unified}. Similar case can be found in \cite{Heemels2009networked}, where there are two zoom parameters (the one is in the coder side and the other is in the decoder side) being updated at the arrival times.
\hfill $\square$
\end{remark}

Since there are some errors induced by the quantizer and the network, the following assumption is used to guarantee that the quantizer does not saturate; see \cite{Heemels2009networked, Nesic2009unified}.

\begin{assumption}
\label{asn-3}
The bound of the initial state $(x_{\p}(t_{0}), x_{\rf}(t_{0}), x_{\ct}(t_{0}))$ is assumed to be known \emph{a priori}. The quantization parameter $\mu\in\mathbb{R}^{l}_{>0}$ is such that the quantization error $\bm{\epsilon}$ is bounded.
\end{assumption}

Assumption \ref{asn-3} ensures that the system state is in the quantization regions, which further implies that the time-scheduling protocol is Lyapunov uniformly globally exponentially stable \cite{Heemels2010networked, Carnevale2007lyapunov, Nesic2004input}. This assumption is enforced easily for linear systems \cite{Nesic2009unified, Liberzon2003stabilization}, and reasonable due to the extensive study on quantized control in the literature. For instance, the bound of the initial state is obtained by an initial zooming-out stage, where the quantization parameter increases such that the state is captured by the quantization regions; see \cite{Liu2015dynamic, Liberzon2007input}. The bound of the quantization error can be obtained for linear systems in \cite{Liu2015dynamic} and for nonlinear systems in \cite{Franci2010quantised}.

\begin{remark}
\label{rmk-6}
In previous works \cite{Liberzon2007input} without Assumption \ref{asn-3}, the quantization mechanism with zooming-out stage is implemented. The goal of the zooming-out stage is to bound the system state in finite time by increasing the quantization parameter. However, since the quantization errors are increasing in the zooming-out stage, the time-scheduling protocol is not Lyapunov uniformly globally asymptotically stable (UGAS); see \cite{Heemels2010networked, Carnevale2007lyapunov, Nesic2004input}. As a result, the stability analysis in this paper can not be applied directly in this case.
\hfill $\square$
\end{remark}

\section{Development of System Model}
\label{sec-developandreformulate}

According to the analysis for information transmission in Subsection \ref{subsec-informationtransmission}, we construct an impulsive model and further a unified hybrid model for the tracking control problem of NQCSs in this section. To this end, the objective of this paper is first transformed to establish the convergence of $x_{\p}$ towards $x_{\rf}$ in the presence of the quantizer and network. To measure the convergence of $x_{\p}$ towards $x_{\rf}$, define the tracking error $\eta:=x_{\p}-x_{\rf}\in\mathbb{R}^{n_{\p}}$ and the error $e_{1}:=(e_{\df}, e_{\ct})\in\mathbb{R}^{n_{1}}$, where $n_{1}=n_{y}+n_{\ct}$. Combining all the variables and analyses in Subsection \ref{subsec-informationtransmission}, the resulting system model, denoted by $\mathcal{S}_{1}$, is presented as the following impulsive system.
\begin{align*}
&\left.\begin{aligned}
\dot{\eta}&=F_{\eta}(\eta, x_{\ct}, x_{\rf}, e_{1}, e_{\f})\\
\dot{x}_{\ct}&=F_{\ct}(\eta, x_{\ct}, x_{\rf}, e_{1}, e_{\f}) \\
\dot{x}_{\rf}&=F_{\rf}(\eta, x_{\ct}, x_{\rf}, e_{1}, e_{\f})  \\
\dot{\mu}&=G_{\mu}(\eta, x_{\ct}, x_{\rf}, e_{1}, e_{\f}, \mu)\\
\dot{e}_{1}&=G_{1}(\eta, x_{\ct}, x_{\rf}, e_{1}, e_{\f}) \\
\dot{e}_{\f}&=G_{\f}(\eta, x_{\ct}, x_{\rf}, e_{1}, e_{\f})
\end{aligned}\right\} t\in[t_{s_{i}}, t_{s_{i+1}}]\setminus\{r_{i}\},\\
&\mu(r^{+}_{s_{i}})=\mu(r_{s_{i}})-\mu(t_{s_{i}})+H_{\mu}(i, \eta(t_{s_{i}}), x_{\ct}(t_{s_{i}}), x_{\rf}(t_{s_{i}}), \\
&\qquad\qquad e_{\df}(t_{s_{i}}), e_{\f}(t_{s_{i}}), \mu(t_{s_{i}})),\\
&e_{1}(r^{+}_{i})=e_{1}(r_{i})-e_{1}(t_{s_{i}})+H_{1}(i, \eta(t_{s_{i}}), x_{\ct}(t_{s_{i}}), x_{\rf}(t_{s_{i}}), \\
&\qquad\qquad e_{1}(t_{s_{i}}), e_{\f}(t_{s_{i}}), \mu(t_{s_{i}})),\\
&e_{\f}(r^{+}_{i})=e_{\f}(r_{i})-e_{\f}(t_{s_{i}})+H_{\f}(i, \eta(t_{s_{i}}), x_{\ct}(t_{s_{i}}), x_{\rf}(t_{s_{i}}), \\
&\qquad\qquad e_{\df}(t_{s_{i}}), e_{\f}(t_{s_{i}}), \mu(t_{s_{i}})).
\end{align*}
All the functions in $\mathcal{S}_{1}$ are derived by detailed calculations and assumed to be continuous. See Appendix for the detailed expressions of all the functions appear in $\mathcal{S}_{1}$.

\begin{remark}
\label{rmk-7}
The impulsive model $\mathcal{S}_{1}$ is general enough to include the one in \cite{Postoyan2014tracking}. If the feedback controller \eqref{eqn-9} is applied, then a similar impulsive model can be obtained along the same fashion. In addition, if the reference system \eqref{eqn-3} is removed, then $\mathcal{S}_{1}$ is reduced to a general model for NQCSs with communication delays. In this case, $\mathcal{S}_{1}$ extends those developed in \cite{Nesic2009unified, Heemels2010networked} by combining all the network-induced issues and casts a new light on stability analysis for NQCSs with communication delays.
\hfill $\square$
\end{remark}

Now, our objective is to establish sufficient conditions to guarantee ISS of the system $\mathcal{S}_{1}$ from $e_{\f}$ to $(\eta, e_{1}, \mu)$. Here, $e_{\f}$ is the network-induced errors, and not necessarily vanishing with the time line; see \cite{Van2010tracking, Postoyan2014tracking}. For specific quantizers and time-scheduling protocols, the effects of $e_{\f}$ can be attenuated, which will be discussed in Section \ref{sec-lyapunovfunction}.

\subsection{Reformulation of System Model}
\label{subsec-reform}

To facilitate stability analysis, the system model $\mathcal{S}_{1}$ is further transformed into a hybrid system by the similar mechanism proposed in \cite{Goebel2006solutions} and employed in \cite{Heemels2010networked, Heemels2009networked, Carnevale2007lyapunov, Postoyan2014tracking}. To this end, some auxiliary variables are introduced. For the sake of convenience, define the augmented state $x:=(\eta, x_{\ct}, x_{\rf})\in\mathbb{R}^{n_{x}}$ and the augmented error $e:=(e_{1}, e_{\f})\in\mathbb{R}^{n_{e}}$, where $n_{x}=n_{\p}+n_{\ct}+n_{\rf}$ and $n_{e}=n_{1}+n_{\f}$. For the update at the arrival times $r_{i}$, $i\in\mathbb{N}$, the variables $m_{1}\in\mathbb{R}^{n_{e}}$ and $m_{2}\in\mathbb{R}^{l}$ are used to store the information $H_{e}(i, x(t_{s_{i}}), e(t_{s_{i}}), \mu(t_{s_{i}}))-e(t_{s_{i}})$ and $H_{\mu}(i, x(t_{s_{i}}), e(t_{s_{i}}), \mu(t_{s_{i}}))-\mu(t_{s_{i}})$, respectively. In addition, denote $m_{1}:=(m^{1}_{1}, m^{\f}_{1})$ and $m^{1}_{1}:=(m^{\df}_{1}, m^{\ct}_{1})$. The variable $\tau$ is a timer to compute both transmission intervals and transmission delays, and $c$ is a variable to count the transmission event. The variable $b$ is a logical variable to show whether the next event is a transmission event or an update event. That is, the existence of $b$ is to make sure that the update event is prior to the next sampling.

Based on above auxiliary variables, the resulting hybrid system, denoted by $\mathcal{S}_{2}$, is presented below. The hybrid system $\mathcal{S}_{2}$ has two part. The first part is flow equation, which is given by
\begin{align}
\label{eqn-14}
\left.\begin{aligned}
\dot{x}&=f(x, e) \\
\dot{e}&=g_{e}(x, e) \\
\dot{\mu}&=g_{\mu}(x, e, \mu)\\
\dot{m}_{1}&=0, \quad \dot{m}_{2}=0 \\
\dot{\tau}&=1, \quad \dot{c}=0, \quad \dot{b}=0
\end{aligned}\right\} &\begin{aligned}
&(b=0\wedge\tau\in[0, h_{\mati}])\\
& \vee(b=1\wedge\tau\in[0, h_{\mad}]).
\end{aligned}
\end{align}
To simplify the notation, denote $\xi:=(x, e, \mu, m_{1}, m_{2})\in\mathbb{R}^{n_{\xi}}$ and $n_{\xi}=n_{x}+2n_{e}+2l$. The second part is the jump equation: $(\xi^{+}, \tau^{+}, c^{+}, b^{+})=R(\xi, \tau, c, b)$. The jump equation also contains two parts for different time instants: the transmission jump equation for the transmission instants $t_{s_{i}}$, $i\in\mathbb{N}$, is given by
\begin{align}
\label{eqn-15}
R(\xi, \tau, c, 0)&=(x, e, \mu, H_{e}(c, x, e, \mu)-e, \nonumber\\
&\quad H_{\mu}(c, x, e, \mu)-\mu, 0, c+1, 1);
\end{align}
and the update jump equation for the arrival instants $r_{i}$, $i\in\mathbb{N}$, is given by
\begin{align}
\label{eqn-16}
R(\xi, \tau, c, 1)&=(x, e+m_{1}, \mu+m_{2}, -e-m_{1}, \nonumber\\
&\quad -\mu-m_{2}, \tau, c, 0).
\end{align}
Note that for the jump equations \eqref{eqn-15}-\eqref{eqn-16}, we have that $(b=0\wedge\tau\in[\varepsilon, h_{\mati}])\vee(b=1\wedge\tau\in[0, h_{\mad}])$ holds, where the constant $\varepsilon>0$ is given in Assumption \ref{asn-2}.

\subsection{Hybrid Model of NQCSs}
\label{subsec-hybrid}

In view of Subsection \ref{subsec-reform}, we further write the system $\mathcal{S}_{2}$ into the formalism as in \cite{Cai2009characterizations, Goebel2006solutions}. Denote $\mathfrak{X}:=(\xi, \tau, c, b)\in\mathscr{R}:=\mathbb{R}^{n_{\xi}}\times\mathbb{R}_{\geq0}\times\mathbb{N}\times\{0, 1\}$. The hybrid system $\mathcal{S}_{2}$ can be rewritten as
\begin{align}
\label{eqn-17}
\mathcal{H}: \left\{\begin{aligned}
\dot{\mathfrak{X}}&=F(\mathfrak{X}), \quad \mathfrak{X}\in C; \\
\mathfrak{X}^{+}&=G(\mathfrak{X}), \quad \mathfrak{X}\in D,
\end{aligned}\right.
\end{align}
where $C:=\{\mathfrak{X}\in\mathscr{R}|(b=0\wedge\tau\in[0, h_{\mati}])\vee(b=1\wedge\tau\in[0, h_{\mad}])\}$ and $D:=\{\mathfrak{X}\in\mathscr{R}|(b=0\wedge\tau\in[\varepsilon, h_{\mati}])\vee(b=1\wedge\tau\in[0, h_{\mad}])\}$. The mapping $F$ in \eqref{eqn-17} is defined as
\begin{align}
\label{eqn-18}
F(\mathfrak{X})&:=(f(x, e), g_{e}(x, e), g_{\mu}(x, e, \mu), 0, 0, 1, 0, 0);
\end{align}
and the mapping $G$ in \eqref{eqn-17} is defined as
\begin{align}
\label{eqn-19}
G(\mathfrak{X})&:=\left\{\begin{aligned}
G_{1}(\mathfrak{X}), \quad \mathfrak{X}\in D_{1}; \\
G_{2}(\mathfrak{X}), \quad \mathfrak{X}\in D_{2},
\end{aligned}\right.
\end{align}
where $G_{1}(\mathfrak{X}):=(x, e, \mu, H_{e}(c, x, e, \mu)-e, H_{\mu}(c, x, e, \mu)-\mu, 0, c+1, 1)$ with $D_{1}:=\{\mathfrak{X}\in\mathscr{R}|b=0\wedge\tau\in[\varepsilon, h_{\mati}]\}$ corresponds to a transmission jump \eqref{eqn-15}; and $G_{2}(\mathfrak{X}):=(x, e+m_{1}, \mu+m_{2}, -e-m_{1}, -\mu-m_{2}, \tau, c, 0)$ with $D_{2}:=\{\mathfrak{X}\in\mathscr{R}|b=1\wedge\tau\in[0, h_{\mad}]\}$ corresponds to the update jump \eqref{eqn-16}.

For the hybrid model $\mathcal{H}$, the sets $C$ and $D$ in \eqref{eqn-17} are closed, and the flow map $F$ in \eqref{eqn-18} is continuous due to the continuity assumptions on $f, g_{e}$ and $g_{\mu}$. The jump map $G$ in \eqref{eqn-19} is continuous and locally bounded by continuity of $G_{1}$ and $G_{2}$. As a result, the hybrid model $\mathcal{H}$ satisfies the basic assumptions given in Section \ref{sec-preliminaries}; see also \cite{Wang2015emulation, Cai2009characterizations, Goebel2006solutions}.

\section{Stability Analysis}
\label{sec-mainresults}

In this section, the main results are presented. Sufficient conditions are derived to guarantee ISS from $e_{\f}$ to $(x, e, \mu)$. Before we state the main results, some assumptions are required. For the $(e, \mu)$-subsystem in flow equation \eqref{eqn-18} and the jump equation \eqref{eqn-19}, the following two assumptions are satisfied.

\begin{assumption}
\label{asn-4}
There exist a function $W: \mathbb{R}^{n_{e}}\times\mathbb{R}^{l}\times\mathbb{R}^{n_{e}}\times\mathbb{R}^{l}\times\mathbb{N}\times\{0, 1\}\rightarrow\mathbb{R}_{>0}$ which is locally Lipschitz in $(e, \mu, m_{1}, m_{2})$ for all $(c, b)\in\mathbb{N}\times\{0, 1\}$, $\alpha_{1W}, \alpha_{2W}, \alpha_{3W}, \alpha_{4W}\in\mathcal{K}_{\infty}$ and $\lambda\in[0, 1)$ such that for all $(e, \mu, m_{1}, m_{2}, c, b)\in\mathbb{R}^{n_{e}}\times\mathbb{R}^{l}\times\mathbb{R}^{n_{e}}\times\mathbb{R}^{l}
\times\mathbb{N}\times\{0, 1\}$,
\begin{align}
\label{eqn-20}
&W(e, \mu, m_{1}, m_{2}, c, b)\geq \alpha_{1W}(|(e, \mu, m_{1}, m_{2})|),\\
\label{eqn-21}
&W(e, \mu, m_{1}, m_{2}, c, b)\leq\alpha_{2W}(|(e, \mu, m_{1}, m_{2})|), \\
\label{eqn-22}
&W(e, \mu, h_{e}(c, x, e, \mu)-e, h_{\mu}(c, x, e, \mu)-\mu, c+1, 1)\nonumber\\
&\ \leq\lambda W(e, \mu, m_{1}, m_{2}, c, 0)+\alpha_{3W}(|e_{\f}|),\\
\label{eqn-23}
&W(e+m_{1}, \mu+m_{2}, -e-m_{1}, -\mu-m_{2}, c, 0)\nonumber\\
&\ \leq W(e, \mu, m_{1}, m_{2}, c, 1)+\alpha_{4W}(|e_{\f}|).
\end{align}
\end{assumption}

\begin{assumption}
\label{asn-5}
For each $b\in\{0, 1\}$, there exist a continuous function $H_{b}: \mathbb{R}^{n_{x}}\rightarrow\mathbb{R}_{>0}$, $\sigma_{bW}\in\mathcal{K}_{\infty}$ and $L_{b}\in[0, \infty)$ such that for all $(x, m_{1}, m_{2}, c, b)\in\mathbb{R}^{n_{x}}\times\mathbb{R}^{n_{e}}\times\mathbb{R}^{l}\times\mathbb{N}\times\{0, 1\}$ and almost all $e\in\mathbb{R}^{n_{e}}, \mu\in\mathbb{R}^{l}$,
\begin{align}
\label{eqn-24}
\mathcal{I}(b)&:=\left\langle\frac{\partial W(e, \mu, m_{1}, m_{2}, c, b)}{\partial e}, g_{e}(x, e)\right\rangle \nonumber \\
&\quad +\left\langle\frac{\partial W(e, \mu, m_{1}, m_{2}, c, b)}{\partial\mu}, g_{\mu}(x, e, \mu)\right\rangle \nonumber \\
&\leq L_{b}W(e, \mu, m_{1}, m_{2}, c, b)+H_{b}(x)+\sigma_{bW}(|e_{\f}|).
\end{align}
\end{assumption}

In Assumptions \ref{asn-4}-\ref{asn-5}, the function $W$ is used to analyze the stability of $(e, \mu)$-subsystem. Assumption \ref{asn-4} is to estimate the jumps of $W$ at the discrete-time instants, i.e., the transmission times and the arrival times. Assumption \ref{asn-5} is to estimate the derivative of $W$ in the continuous-time intervals. Since Assumptions \ref{asn-4}-\ref{asn-5} are applied to the $(e, \mu)$-subsystem, \eqref{eqn-22}-\eqref{eqn-24} hold with respect to the additional item $e_{\f}$, which are parts of $e$. As a result, Assumptions \ref{asn-4}-\ref{asn-5} are different from the classic assumptions like Condition \uppercase\expandafter{\romannumeral4}.1 in \cite{Heemels2010networked} and Assumption 1 in \cite{Carnevale2007lyapunov}, where the external disturbances are involved. Moreover, $\alpha_{3W}$ and $\alpha_{4W}$ in Assumption \ref{asn-4} are allowed to be the same. For instance, \eqref{eqn-22}-\eqref{eqn-23} hold with $\bar{\alpha}_{3W}(v):=\bar{\alpha}_{4W}(v):=\max\{\alpha_{3W}(v), \alpha_{4W}(v)\}$, respectively. On the other hand, similar conditions to Assumptions \ref{asn-4}-\ref{asn-5} have been considered in previous works \cite{Tabbara2008input, Postoyan2014tracking}. However, external disturbances have been studied in \cite{Tabbara2008input}, whereas $e_{\f}$ is part of the augmented error $e$ in Assumptions \ref{asn-4}-\ref{asn-5}. Both quantization and transmission delays have not been considered in \cite{Postoyan2014tracking}, and however are studied in this paper. The construction of $W$ satisfying Assumptions \ref{asn-4}-\ref{asn-5} will be presented in Section \ref{sec-lyapunovfunction}.

\begin{assumption}
\label{asn-6}
There exist a locally Lipschitz function $V: \mathbb{R}^{n_{x}}\rightarrow\mathbb{R}_{\geq0}$, $\alpha_{1V}, \alpha_{2V}, \sigma_{bV}\in\mathcal{K}_{\infty}$, and constants $\rho_{b}, \theta_{b}, \gamma_{b}>0$ for $b\in\{0, 1\}$, such that for all $x\in\mathbb{R}^{n_{x}}$,
\begin{align}
\label{eqn-25}
&\alpha_{1V}(|x|)\leq V(x)\leq\alpha_{2V}(|x|),
\end{align}
and for all $(e, \mu, m_{1}, m_{2}, c, b)\in\mathbb{R}^{n_{e}}\times\mathbb{R}^{l}\times\mathbb{R}^{n_{e}}\times\mathbb{R}^{l}
\times\mathbb{N}\times\{0, 1\}$ and almost all $x\in\mathbb{R}^{n_{x}}$,
\begin{align}
\label{eqn-26}
&\langle\nabla V(x), f(x, e)\rangle\leq-\rho_{b}V(x)-H^{2}_{b}(x)+(\gamma^{2}_{b}-\theta_{b})  \nonumber\\
&\quad \times W^{2}(e, \mu, m_{1}, m_{2}, c, b)+\sigma_{bV}(|e_{\f}|),
\end{align}
where $H_{b}: \mathbb{R}^{n_{x}}\rightarrow\mathbb{R}_{>0}$ is defined in Assumption \ref{asn-5}.
\end{assumption}

Assumption \ref{asn-6} is  a robust stability property of the closed-loop system \eqref{eqn-2}-\eqref{eqn-5}, and the function $V$ is used to analyze the stability of the $x$-subsystem; see \cite{Postoyan2014tracking}. Observe from \eqref{eqn-26} that $e_{\f}$ is treated as the external input of the $x$-subsystem, which relaxes the standard assumptions in \cite{Heemels2009networked, Heemels2010networked, Nesic2009unified}. In addition, Assumption \ref{asn-6} implies that the designed controller leads the $\eta$-subsystem (i.e., the tracking error dynamics) to have ISS-like property from $(W, e_{\f})$ to $x$. Moreover, Assumption \ref{asn-6} also indicates that the $x$-subsystem is $\mathcal{L}_{2}$-stable from $(W, e_{\f})$ to $H_{b}$; see \cite[Theorem 4]{Nesic2004input} and \cite[Remark \uppercase\expandafter{\romannumeral5}.2]{Heemels2010networked}. In Section \ref{sec-illustration}, a numerical example is presented to demonstrate how to verify Assumptions \ref{asn-4}-\ref{asn-6}.

Although the transmission intervals and the transmission delays are bounded in Assumption \ref{asn-1}, the tradeoff between $h_{\mati}$ and $h_{\mad}$ needs to be studied further. Consider the following differential equations
\begin{align}
\label{eqn-27}
\dot{\bar{\phi}}_{b}=-2L_{b}\bar{\phi}_{b}-\gamma_{b}[(1+\varrho_{b})\bar{\phi}^{2}_{b}+1],
\end{align}
where $b\in\{0, 1\}$, and $L_{b}\geq0, \gamma_{b}>0$ are given in Assumptions \ref{asn-5}-\ref{asn-6}. In addition, $\bar{\phi}_{b}(0):=\bar{\phi}_{b0}\in(1, \lambda^{-1})$, and $\varrho_{b}\in(0, \lambda^{-2}\bar{\phi}^{-2}_{b0}-1)$, where $\lambda$ is given in Assumption \ref{asn-4}. Based on Claim 1 in \cite{Carnevale2007lyapunov} and Claim 1 in \cite{Postoyan2014tracking}, the solutions to \eqref{eqn-26} are strictly decreasing as long as $\bar{\phi}_{b}(\tau)\geq0$, $b\in\{0, 1\}$.

Now we are in the position to state the main result of this section. Its proof is based on the comparison principle \cite[Lemma C.1]{Cai2009characterizations} for hybrid systems and the proof of Theorem IV. 2 in \cite{Heemels2010networked}. However, some essential modifications are required to deal with the effects of the networked-induced errors.

\begin{theorem}
\label{thm-1}
Consider the system $\mathcal{S}_{2}$ and let Assumptions \ref{asn-4}-\ref{asn-6} hold. If the MATI $h_{\mati}$ and the MAD $h_{\mad}$ satisfy
\begin{subequations}
\label{eqn-28}
\begin{align}
\label{eqn-28-1}
\gamma_{0}\bar{\phi}_{0}(\tau)&\geq(1+\varrho_{1})\lambda^{2}\gamma_{1}\bar{\phi}_{1}(0), \quad \tau\in[0, h_{\mati}],\\
\label{eqn-28-2}
\gamma_{1}\bar{\phi}_{1}(\tau)&\geq(1+\varrho_{0})\gamma_{0}\bar{\phi}_{0}(\tau), \quad \tau\in[0, h_{\mad}],
\end{align}
\end{subequations}
with $\bar{\phi}_{0}(0)>0$, $\bar{\phi}_{1}(0)>0$, $\bar{\phi}_{0}(h_{\mati})>0$, then the system $\mathcal{S}_{2}$ is ISS from $ e_{\f}$ to $(\eta, e_{1}, \mu)$. That is, there exist $\beta\in\mathcal{KLL}$ and $\varphi_{1}\in\mathcal{K}_{\infty}$ such that for all the solution of $\mathcal{S}_{2}$ and all $(t, j)\in\mathbb{R}_{\geq0}\times\mathbb{N}$ in the admissible domain of the solution,
\begin{align}
\label{eqn-29}
&|(x(t, j), e(t, j), \mu(t, j))|\nonumber\\
&\leq\beta(|(x(t_{0}, j_{0}), e(t_{0}, j_{0}), \mu(t_{0}, j_{0}))|, t-t_{0}, j-j_{0})\nonumber\\
&\quad +\varphi_{1}(\|e_{\f}\|_{(t, j)}).
\end{align}
\end{theorem}

\begin{IEEEproof}
Consider the hybrid system $\mathcal{H}$ in \eqref{eqn-17}. For $\mathfrak{X}\in C\cup D\cup G(D)$, define the following Lyapunov function:
\begin{equation}
\label{eqn-30}
U(\mathfrak{X}):=V(x)+\gamma_{b}\bar{\phi}_{b}(\tau)W^{2}(e, \mu, m_{1}, m_{2}, c, b).
\end{equation}
Using \eqref{eqn-20}, \eqref{eqn-21} and \eqref{eqn-25}, it follows that
\begin{align*}
U(\mathfrak{X})&\geq\alpha_{1V}(|x|)+\gamma_{0}\bar{\phi}_{0}(\tau)\alpha^{2}_{1W}(|(e, \mu, m_{1}, m_{2})|),\\
U(\mathfrak{X})&\leq\alpha_{2V}(|x|)+\gamma_{1}\bar{\phi}_{1}(\tau)\alpha^{2}_{2W}(|(e_{1}, \mu, m^{1}_{1}, m_{2})|).
\end{align*}
According to \eqref{eqn-28}, there exist $\alpha_{1}, \alpha_{2}\in\mathcal{K}_{\infty}$ such that
\begin{align}
\label{eqn-31}
\alpha_{1}(|\xi|)\leq U(\xi)\leq\alpha_{2}(|\xi|),
\end{align}
where,
\begin{align*}
\alpha_{1}(v)&:=\min\{\alpha_{1V}(v/2), (1+\varrho_{1})\lambda^{2}\gamma_{1}\bar{\phi}_{1}(0)\alpha^{2}_{1W}(v/2)\}, \\ \alpha_{2}(v)&:=\max\{\alpha_{2V}(v), \gamma_{1}\bar{\phi}_{1}(0)\alpha^{2}_{2W}(v)\}.
\end{align*}

Next, consider the evolution of $U$ in the continuous-time intervals and at the discrete-time instants, respectively. For the flow equation $F$ in \eqref{eqn-18}, we have that\footnote{$\langle\nabla U(\mathfrak{X}), F(\mathfrak{X})\rangle$ is used here with a slight abuse of terminology since $U$ is not differential almost everywhere. However, this is justified by the fact that $\dot{c}=0$ and $\dot{b}=0$ in \eqref{eqn-14}.}
\begin{align}
\label{eqn-32}
&\langle\nabla U(\mathfrak{X}), F(\mathfrak{X})\rangle \nonumber\\
&=\langle\nabla V(x), f(x, e)\rangle+\gamma_{b}\dot{\bar{\phi}}_{b}(\tau)W^{2}(e, \mu, m_{1}, m_{2}, c, b) \nonumber \\
&\quad +2\gamma_{b}\bar{\phi}_{b}(\tau)W(e, \mu, m_{1}, m_{2}, c, b) \nonumber \\
&\quad \times \left(\left\langle\frac{\partial W(e, \mu, m_{1}, m_{2}, c, b)}{\partial e}, g_{e}(x, e)\right\rangle\right.  \nonumber \\
&\quad \left.+\left\langle\frac{\partial W(e, \mu, m_{1}, m_{2}, c, b)}{\partial\mu}, g_{\mu}(x, e, \mu)\right\rangle\right) \nonumber \\
&\leq-\rho_{b}V(x)-\theta_{b}W^{2}(e, \mu, m_{1}, m_{2}, c, b)-H^{2}_{b}(x) \nonumber\\
&\quad+\gamma^{2}_{b}W^{2}(e, \mu, m_{1}, m_{2}, c, b)+ \sigma_{bV}(|e_{\f}|) \nonumber\\
&\quad+2\gamma_{b}\bar{\phi}_{b}(\tau)W(e, \mu, m_{1}, m_{2}, c, b) \nonumber\\
&\quad\times[L_{b}W(e, \mu, m_{1}, m_{2}, c, b)+H_{b}(x)+\sigma_{bW}(|e_{\f}|)] \nonumber\\
&\quad+W^{2}(e, \mu, m_{1}, m_{2}, c, b) \nonumber\\
&\quad\times\gamma_{b}[-2L_{b}\bar{\phi}_{b}(\tau)-\gamma_{b}((1+\varrho_{b})\bar{\phi}^{2}_{b}(\tau)+1)]\nonumber\\
&=-\rho_{b}V(x)-\theta_{b}W^{2}(e, \mu, m_{1}, m_{2}, c, b)-H^{2}_{b}(x) \nonumber\\
&\quad+\sigma_{bV}(|e_{\f}|)+2\gamma_{b}\bar{\phi}_{b}(\tau)W(e, \mu, m_{1}, m_{2}, c, b)[H_{b}(x) \nonumber\\
&\quad+\sigma_{bW}(|e_{\f}|)]-(1+\varrho_{b})\gamma^{2}_{b}\bar{\phi}^{2}_{b}(\tau)W^{2}(e, \mu, m_{1}, m_{2}, c, b) \nonumber\\
&\leq-\rho_{b}V(x)-\theta_{b}W^{2}(e, \mu, m_{1}, m_{2}, c, b)-[H_{b}(x) \nonumber\\
&\quad-\gamma_{b}\bar{\phi}_{b}(\tau)W(e, \mu, m_{1}, m_{2}, c, b)]^{2}+\bar{\sigma}_{b}(|e_{\f}|) \nonumber\\
&\leq-\rho_{b}V(x)-\theta_{b}W^{2}(e, \mu, m_{1}, m_{2}, c, b)+\bar{\sigma}_{b}(|e_{\f}|),
\end{align}
where, the first ``='' holds due to the definition of the flow equation; the first ``$\leq$'' holds because of the functions in \eqref{eqn-27} and Assumptions \ref{asn-5}-\ref{asn-6}; the second ``$\leq$'' holds due to the fact that $2xy\leq\ell x^{2}+y^{2}/\ell$ for all $x, y\geq0$ and $\ell>0$. In addition, $\bar{\sigma}_{b}(v):=\sigma^{2}_{bW}(v)/\varrho_{b}+\sigma_{bV}(v)$ with $b\in\{0, 1\}$.

For the jump equation $G$ in \eqref{eqn-19}, there are two cases based on the values of the variable $b$. For the case $b=0$,
\begin{align}
\label{eqn-33}
U(G_{1}(\mathfrak{X}))&=V(x)+\gamma_{1}\bar{\phi}_{1}(0)W^{2}(e, \mu, h_{e}(c, x, e, \mu)-e,\nonumber\\
&\qquad h_{\mu}(c, x, e, \mu)-\mu, c+1, 1)\nonumber \\
&\leq V(x)+\gamma_{1}\bar{\phi}_{1}(0)[\lambda W(e, \mu, m_{1}, m_{2}, c, 0)+\alpha_{3W}(|e_{\f}|)]^{2}\nonumber\\
&\leq V(x)+\gamma_{1}\bar{\phi}_{1}(0)[(1+\varrho_{1})\lambda^{2}W^{2}(e, a, \mu, c, 0)\nonumber\\
&\quad\left.+\left(1+\frac{1}{\varrho_{1}}\right)\alpha^{2}_{3W}(|e_{\f}|)\right]\nonumber\\
&\leq V(x)+\gamma_{0}\bar{\phi}_{0}(\tau)W^{2}(e, \mu, m_{1}, m_{2}, c, 0)+\alpha_{31}(|e_{\f}|)\nonumber\\
&=U(\mathfrak{X})+\alpha_{31}(|e_{\f}|),
\end{align}
where, the first ``='' holds due to the definition of the jump equation; the first ``$\leq$'' holds due to Assumption \ref{asn-4}; the second ``$\leq$'' holds due to the fact that $2xy\leq\ell x^{2}+y^{2}/\ell$ for all $x, y\geq0$ and $\ell>0$; the third ``$\leq$'' holds due to the inequality \eqref{eqn-28-1}. In addition, $\alpha_{31}(v):=(\varrho_{1}\lambda^{2})^{-1}\gamma_{0}\bar{\phi}_{0}(\tau)\alpha^{2}_{3W}(v)$. Similarly, for the case $b=1$,
\begin{align}
\label{eqn-34}
U(G_{2}(\mathfrak{X}))&=V(x)+\gamma_{0}\bar{\phi}_{0}(\tau)W^{2}(e+m_{1}, \mu+m_{2}, \nonumber\\
&\quad-e-m_{1},\mu-m_{2}, c, 0)\nonumber\\
&\leq V(x)+\gamma_{0}\bar{\phi}_{0}(\tau)[W(e, \mu, m_{1}, m_{2}, c, 1)  \nonumber\\
&\quad +\alpha_{4W}(|e_{\f}|)]^{2}\nonumber\\
&\leq V(x)+\gamma_{0}\bar{\phi}_{0}(\tau)[(1+\varrho_{0})W^{2}(e, \mu, m_{1}, m_{2}, c, 1)\nonumber\\
&\quad\left.+\left(1+\frac{1}{\varrho_{0}}\right)\alpha^{2}_{4W}(|e_{\f}|)\right]\nonumber\\
&\leq V(x)+(1+\varrho_{0})\gamma_{0}\bar{\phi}_{0}(\tau)W^{2}(e, \mu, m_{1}, m_{2}, c, 1) \nonumber\\
&\quad +\alpha_{32}(|e_{\f}|)\nonumber\\
&=U(\mathfrak{X})+\alpha_{32}(|e_{\f}|),
\end{align}
where $\alpha_{32}(v):=\varrho_{0}^{-1}\gamma_{0}\bar{\phi}_{0}(\tau)\alpha^{2}_{4W}(v)$.

Combining \eqref{eqn-32}-\eqref{eqn-34}, we obtain that
\begin{align*}
\langle\nabla U(\mathfrak{X}), F(\mathfrak{X})\rangle&\leq-\rho_{b}V(x)-\theta_{b}W^{2}(e, \mu, m_{1}, m_{2}, c, b) \\
&\quad +\bar{\sigma}_{b}(|e_{\f}|),\\
U(G(\mathfrak{X}))&\leq U(\mathfrak{X})+\alpha_{3}(|e_{\f}|).
\end{align*}
Define $\bar{\varepsilon}:=\min\{\rho_{b}, \theta_{b}\}$ and $\tilde{\varepsilon}\in(0, \bar{\varepsilon}\min\{1, \gamma^{-1}_{b}\bar{\phi}_{b}(0)\})$. We obtain by calculation that
\begin{align}
\label{eqn-35}
\langle\nabla U(\mathfrak{X}), F(\mathfrak{X})\rangle&\leq-\tilde{\varepsilon}U(\mathfrak{X})+\sigma_{1}(|e_{\f}|),\\
\label{eqn-36}
U(\mathfrak{X}(t_{i}, j+1))&\leq U(\mathfrak{X}(t_{i}, j))+\alpha_{3}(|e_{\f}|),
\end{align}
where $\sigma_{1}(v):=\max\{\bar{\sigma}_{1}(v), \bar{\sigma}_{2}(v)\}$ and $\alpha_{3}(v):=\max\{\alpha_{31}(v), \alpha_{32}(v)\}$.

Integrating and iterating \eqref{eqn-35}-\eqref{eqn-36} from $(t_{0}, j_{0})$ to $(t, j)$ in the hybrid time domain, one has
\begin{align}
\label{eqn-37}
U(\mathfrak{X}(t, j))&\leq e^{-\tilde{\varepsilon}(t-t_{0})}U(\mathfrak{X}(t_{0}, j_{0}))  \nonumber\\
&\quad +\frac{1}{1-e^{-\varepsilon\tilde{\varepsilon}}}[\tilde{\varepsilon}^{-1}\sigma_{1}(\|e_{\f}\|_{(t, j)})+\alpha_{3}(\|e_{\f}\|_{(t, j)})],
\end{align}
where $\varepsilon>0$ is given in Assumption \ref{asn-1}. From \eqref{eqn-31} and \eqref{eqn-37}, we have
\begin{align*}
&|(\eta(t, j), e_{1}(t, j), \mu(t, j))|\\
&\leq\alpha^{-1}_{1}(2e^{-\tilde{\varepsilon}(t-t_{0})}\alpha_{2}(|\mathfrak{X}(t_{0}, j_{0})|))\\
&\quad +\alpha^{-1}_{1}\left(\frac{4}{1-e^{-\varepsilon\tilde{\varepsilon}}}[\tilde{\varepsilon}^{-1}\sigma_{1}(\|e_{\f}\|_{(t, j)})+\alpha_{3}(\|e_{\f}\|_{(t, j)})]\right).
\end{align*}
Thus, the system $\mathcal{S}_{2}$ is ISS from $(e_{\f}, w)$ to $(\eta, e_{1}, \mu)$ with
\begin{align*}
\beta(v, t, j)&:=\alpha^{-1}_{1}(2e^{-\tilde{\varepsilon}(0.5t+0.5\varepsilon j)}\alpha_{2}(v)) \\
\varphi_{1}(v)&:=\alpha^{-1}_{1}\left(\frac{4(\tilde{\varepsilon}^{-1}\sigma_{1}(v)+\alpha_{3}(v))}{1-e^{-\varepsilon\tilde{\varepsilon}}}\right),
\end{align*}
where the definition of $\beta$ comes from the fact that $t\geq\varepsilon j$ (see also \cite[Section V-A]{Carnevale2007lyapunov}), and $\varepsilon>0$ is given in Assumption \ref{asn-1}. As a result, the proof is completed.
\end{IEEEproof}

\begin{remark}
\label{rmk-8}
In Theorem \ref{thm-1}, \eqref{eqn-28-1}-\eqref{eqn-28-2} for the MATI $h_{\mati}$ and the MAD $h_{\mad}$ are different from those in \cite{Carnevale2007lyapunov, Postoyan2014tracking, Heemels2010networked}. For the delay-free and quantization-free case, the formula of the MATI was given explicitly in \cite[Assumption 4]{Postoyan2014tracking}. In fact, the MATI in \cite{Postoyan2014tracking} is the same as the one in \cite{Carnevale2007lyapunov}, and is the solution to the equation $\dot{\bar{\phi}}=-2L\bar{\phi}-\gamma(\bar{\phi}^{2}+1)$, where $L, \gamma>0$. Such equation has the similar form as \eqref{eqn-27}. As a result, the delay-free and quantization-free case in \cite{Postoyan2014tracking} is recovered as a particular case of this paper.
\hfill $\square$
\end{remark}

Note that if $W$ and $V$ are respectively lower bounded by the functions of $(e, \mu)$ and $x$, then the system $\mathcal{S}_{2}$ is ISS from $e_{\f}$ to $(x, e, \mu)$ along the similar line as in the proof of Theorem \ref{thm-1}. In addition, according to Theorem \ref{thm-1}, the following corollary is an direct consequence, and hence the proof is omitted here.

\begin{corollary}
\label{cor-1}
If all the assumptions in Theorem \ref{thm-1} are satisfied with $\alpha_{1V}(v)=a_{1V}v^{2}$, $\alpha_{2V}(v)=a_{2V}v^{2}$, $\alpha_{1W}(v)=a_{1W}v^{2}$ and $\alpha_{2W}(v)=a_{2W}v^{2}$, where $a_{1V}, a_{2V}, a_{1W}, a_{2W}>0$, then the system $\mathcal{S}_{2}$ is EISS from $e_{\f}$ to $(\eta, e_{1}, \mu)$.
\end{corollary}

In Theorem \ref{thm-1}, the tracking error is established to converge to the region around of the origin. The radius of such a region is related to the ISS nonlinear gain $\varphi_{1}$, which is the function of the norm of $e_{\f}$, respectively. In the following, we further study the ISS gains obtained in Theorem \ref{thm-1}, and develop the relation among the ISS nonlinear gain $\varphi_{1}(v)$ in \eqref{eqn-29}, the MATI $h_{\mati}$, and the MAD $h_{\mad}$.

\begin{proposition}
\label{prop-1}
If all the assumptions in Theorem \ref{thm-1} are satisfied, then $\varphi_{1}(v)$ in \eqref{eqn-29} can be written as the form $(1+\bar{\varphi}(h_{\mati}, h_{\mad}))\hat{\varphi}(\varepsilon)\tilde{\varphi}(v)$, where $\bar{\varphi}: \mathbb{R}_{\geq0}\times\mathbb{R}_{\geq0}\rightarrow\mathbb{R}_{\geq0}$ is a positive definitive function, $\hat{\varphi}: \mathbb{R}_{\geq0}\rightarrow\mathbb{R}_{\geq0}$ is a strictly decreasing function and $\tilde{\varphi}\in\mathcal{K}_{\infty}$.
\end{proposition}

\begin{IEEEproof}
Following the same line as in the proof of Theorem \ref{thm-1}, we have $\varphi_{1}(v)$ with the following form
\begin{align*}
\varphi_{2}(v)&=\alpha^{-1}_{1}\left(\frac{4(\tilde{\varepsilon}^{-1}\sigma_{1}(v)+\alpha_{3}(v))}{1-e^{-\varepsilon\tilde{\varepsilon}}}\right).
\end{align*}

Since $\alpha_{1}(v)=\min\{\alpha_{1V}(v/2), (1+\varrho_{1})\bar{\phi}^{-1}_{b0}\alpha^{2}_{1W}(v/2)\}$ in \eqref{eqn-31}, we obtain that $\alpha_{1}(v)\geq\tilde{\alpha}_{1}(v):=\min\{\alpha_{1V}(v/2), (1+\varrho_{1})\lambda\alpha^{2}_{1W}(v/2)\}$. For $b\in\{0, 1\}$, define the constant $\varrho^{-1}_{b}:=\psi_{b}(h_{\mati}, h_{\mad})$ with certain positive definitive functions $\psi_{b}:\mathbb{R}_{\geq0}\times\mathbb{R}_{\geq0}\rightarrow\mathbb{R}_{\geq0}$. It holds from the definitions of the functions $\sigma_{1}$ and $\alpha_{3}$ in \eqref{eqn-35}-\eqref{eqn-36} that
\begin{align}
\label{eqn-38}
\varphi_{1}(v)&\leq\tilde{\alpha}^{-1}_{1}\left(\frac{4}{1-e^{-\varepsilon\tilde{\varepsilon}}}[\tilde{\varepsilon}^{-1}
\sigma_{1}(v)+\alpha_{3}(v)]\right)\nonumber\\
&\leq\tilde{\alpha}^{-1}_{1}\left(\frac{4}{1-e^{-\varepsilon\tilde{\varepsilon}}}\left[\tilde{\varepsilon}^{-1}\max_{b\in\{0, 1\}}
\left\{\frac{\sigma^{2}_{b1W}(v)}{\varrho_{b}}+\sigma_{b1V}(v)\right\}\right.\right.\nonumber\\
&\quad +\left.\left.\max\left\{\frac{\gamma_{0}\bar{\phi}_{0}(\tau)}{\varrho_{1}\lambda^{2}}\alpha^{2}_{3W}(v), \frac{\gamma_{1}\bar{\phi}_{1}(\tau)}{\varrho_{0}}\alpha^{2}_{4W}(v)\right\}\right]\right)\nonumber\\
&\leq\tilde{\alpha}^{-1}_{1}\left(\frac{4}{1-e^{-\varepsilon\bar{\varepsilon}\min\{1, \lambda/\gamma_{b}\}}}\left[\frac{1}{\bar{\varepsilon}\min\{1, \gamma^{-1}_{b}\lambda\}}\right.\right.\nonumber\\
&\quad \times\max_{b\in\{0, 1\}}\left\{\psi_{b}(h_{\mati}, h_{\mad})\sigma^{2}_{bW}(v)+\sigma_{bV}(v)\right\}\nonumber\\
&\quad +\max_{b\in\{0, 1\}}\left\{\gamma_{b}\bar{\phi}_{b}(0)\psi_{b}(h_{\mati}, h_{\mad})\right\}\nonumber\\
&\quad \left.\left.\times(\lambda^{-2}\alpha^{2}_{3W}(v)+\alpha^{2}_{4W}(v))\right]\right)\nonumber\\
&\leq\hat{\alpha}_{1}\left(\frac{4}{1-e^{-\varepsilon\bar{\varepsilon}\min\{1, \lambda/\gamma_{b}\}}}\right)\left[
\hat{\alpha}_{2}\left(\frac{2}{\bar{\varepsilon}\min\{1, \gamma^{-1}_{b}\lambda\}}\right.\right.\nonumber\\
&\quad\times\max_{b\in\{0, 1\}}\{\psi_{b}(h_{\mati}, h_{\mad})\}\sigma^{2}_{bW}(v))\nonumber\\
&\quad+\hat{\alpha}_{2}\left(\frac{4\sigma_{bV}(v)}{\bar{\varepsilon}\min\{1, \gamma^{-1}_{b}\lambda\}}\right)\nonumber\\
&\quad+\hat{\alpha}_{2}\left(\max_{b\in\{0, 1\}}\{4\gamma_{b}\bar{\phi}_{b}(0)\psi_{b}(h_{\mati}, h_{\mad})\}\right.\nonumber\\
&\quad\left.\left.\times(\lambda^{-2}\alpha^{2}_{3W}(v)+\alpha^{2}_{4W}(v))\right)\right],
\end{align}
where $\hat{\alpha}_{1}, \hat{\alpha}_{2}\in\mathcal{K}_{\infty}$ and the fourth ``$\leq$'' holds because of the following two facts \cite{Kellett2014compendium}: for all $\alpha\in\mathcal{K}_{\infty}$ and $x, y\geq0$, (i) $\alpha(x+y)\leq\alpha(2x)+\alpha(2y)$; (ii) there exist $\alpha_{1}, \alpha_{2}\in\mathcal{K}_{\infty}$ such that $\alpha(xy)\leq\alpha_{1}(x)\alpha_{2}(y)$. Using such two facts several times, there exist a strictly decreasing function $\hat{\varphi}_{1}: \mathbb{R}_{\geq0}\rightarrow\mathbb{R}_{\geq0}$, a positive definitive function $\bar{\varphi}_{1}: \mathbb{R}_{\geq0}\times\mathbb{R}_{\geq0}\rightarrow\mathbb{R}_{\geq0}$ and $\tilde{\varphi}_{1}\in\mathcal{K}_{\infty}$ such that $\varphi_{1}(v)\leq\hat{\varphi}_{1}(\varepsilon)(1+\bar{\varphi}_{1}(h_{\mati}, h_{\mad}))\tilde{\varphi}_{1}(v)$. Therefore, the proof is completed.
\end{IEEEproof}

\begin{remark}
\label{rmk-9}
From the proofs of Theorem \ref{thm-1} and Proposition \ref{prop-1}, the norm $\|e_{\f}\|_{(t, j)}$ and the function $\varphi_{1}$ are related to the MATI $h_{\mati}$ and the MAD $h_{\mad}$. Since ZOH devices are applied, the bound on $\|e_{\f}\|_{(t, j)}$ can be established via a step-by-step sampling approach; see \cite[Section 4.1]{Van2010tracking}. Thus, the upper bound on $\tilde{\varphi}_{1}$ can also be obtained. In addition, the function $\hat{\varphi}_{1}$ is related to the minimum inter-transmission interval $\varepsilon$. From \eqref{eqn-37}, $\hat{\varphi}_{1}$ increases to $\infty$ as $\varepsilon\rightarrow0$. That is, the larger $\varepsilon$ is, the smaller $\hat{\varphi}_{1}$ is. Due to the existence of $\varepsilon$ in all practical networks, the effects of $\varepsilon$ on $\hat{\varphi}_{1}$ can not be avoided but be limited by choosing appropriate networks, thereby leading to the minimum of $\hat{\varphi}_{1}$.
\hfill $\square$
\end{remark}

Observe from Theorem \ref{thm-1} and Proposition \ref{prop-1} that the conservatism of the obtained results comes from the ISS gain $\varphi_{1}$. We would like to obtain the ISS gains as small as possible, which in turn implies better tracking performance. If the feedforward input $u_{\f}$ is transmitted to the reference system directly, then $e_{\f}\equiv0$ and $\varphi_{1}\equiv0$. In addition, the Lyapunov function satisfying Assumptions \ref{asn-4}-\ref{asn-5} is constructed in Section \ref{sec-lyapunovfunction}, and $\alpha_{3W}=\alpha_{3W}\equiv0$, which thus reduces the conservatism of the obtained results. On the other hand, if the reference trajectory is bounded, we would like to guarantee the boundedness of $(x, e, \mu)$. For this case, the following corollary is established, which is an extension of Proposition 1 in \cite{Postoyan2014tracking} and thus the proof is omitted here.

\begin{corollary}
\label{cor-2}
If all the assumptions in Theorem \ref{thm-1} are satisfied, and there exist $\gamma_{\ct}\in\mathcal{K}_{\infty}$, $\beta_{\rf}: \mathbb{R}^{n_{\rf}}\times\mathbb{R}^{n_{\f}}\rightarrow\mathbb{R}_{\geq0}$, $\beta_{\ct}: \mathbb{R}^{n_{\ct}}\rightarrow\mathbb{R}_{\geq0}$ such that for all $(t, j)\in\mathbb{R}_{\geq0}\times\mathbb{N}$ in the admissible domain of the solution,
\begin{align}
\label{eqn-39}
&|(x_{\rf}(t, j), e_{\f}(t, j))|\leq\beta_{\rf}(x_{\rf}(t_{0}, j_{0}), e_{\f}(t_{0}, j_{0})), \\
\label{eqn-40}
&|x_{\ct}(t, j)|\leq\beta_{\ct}(x_{\ct}(t_{0}, j_{0}))+\gamma_{\ct}(\|(\eta, x_{\rf}, e_{1})\|_{(t, j)}),
\end{align}
then there exists a function $\bar{\beta}: \mathbb{R}^{n_{x}}\times\mathbb{R}^{n_{e}}\times\mathbb{R}^{l}\rightarrow\mathbb{R}_{\geq0}$ such that
\begin{align*}
|(x(t, j), e(t, j), \mu(t, j))|\leq\bar{\beta}((x(t_{0}, j_{0}), e(t_{0}, j_{0}), \mu(t_{0}, j_{0}))).
\end{align*}
\end{corollary}

In Corollary \ref{cor-2}, the conditions \eqref{eqn-39}-\eqref{eqn-40} hold due to the boundedness of the reference trajectory. In \eqref{eqn-39}, the boundedness of the reference state comes obviously from the bounded reference trajectory. If the feedforward input $u_{\f}$ is bounded, then $e_{\f}$ is also bounded. The boundedness of $x_{\ct}$ is from Assumption \ref{asn-6}, which implies that the designed controller leads to the ISS-like property from $(W, e_{\f})$ to $\eta$; see also \cite[Theorem 4]{Nesic2004input} and \cite[Remark \uppercase\expandafter{\romannumeral5}.2]{Heemels2010networked}. The interested readers are referred to \cite{Postoyan2014tracking} for more details.

\section{On the Existence of Lyapunov Functions}
\label{sec-lyapunovfunction}

In this section, the existence of the Lyapunov function $W$ satisfying Assumptions \ref{asn-4}-\ref{asn-5} is discussed. Since the network-induced errors are involved, which leads to different assumptions on Lyapunov functions, it is necessary to verify the existence of these Lyapunov functions to ensure the satisfaction of the applied assumptions. Therefore, based on the construction of Lyapunov functions in \cite{Nesic2009unified, Heemels2009networked, Heemels2010networked}, Assumptions \ref{asn-4}-\ref{asn-5} are verified under some appropriate conditions. In addition, explicit Lyapunov functions for different time-scheduling protocols and quantizers are presented.

\subsection{Time-Scheduling Protocols}
\label{subsec-protocol}

In Section \ref{sec-problemformation}, the update of $\bm{\vartheta}=(e_{\df}, e_{\ct}, e_{\f})$ is given by
\begin{equation}
\label{eqn-41}
\bm{\vartheta}(r^{+}_{i})=\bm{\vartheta}(r_{i})-\bm{\vartheta}(t_{s_{i}})+H_{\bm{\vartheta}}(i, x(t_{s_{i}}), \bm{\vartheta}(t_{s_{i}}), \mu(t_{s_{i}})),
\end{equation}
where $H_{\bm{\vartheta}}:=(h_{\df}, h_{\ct}, h_{\f})$ is the update function. Similar to the analysis and the terminology in \cite{Heemels2010networked, Heemels2009networked}, the function $H_{\bm{\vartheta}}(i, x(t_{s_{i}}), \bm{\vartheta}(t_{s_{i}}), \mu(t_{s_{i}}))$ is referred to as the protocol. Based on $l$ nodes of the network, $\bm{\vartheta}$ is partitioned into $\bm{\vartheta}=(\bm{\vartheta}_{1}, \ldots, \bm{\vartheta}_{l})$. If the $j$-th node is granted to access to the network according to certain time-scheduling protocol, then the corresponding component $\bm{\vartheta}_{j}$ is updated and the other components are kept. In the literature, there are several time-scheduling protocols that can be modeled as $H_{\bm{\vartheta}}$; see \cite{Heemels2010networked, Heemels2009networked, Nesic2009unified}. In the following, two classes of commonly-used protocols are presented.

The first protocol is Round-Robin (RR) protocol \cite{Nesic2004input}, which is a periodic protocol \cite{Donkers2011stability}. The period of the RR protocol is $l$, and each node has and only has one chance to access to the network in a period. The function $H_{\bm{\vartheta}}$ is given by
\begin{equation}
\label{eqn-42}
H_{\bm{\vartheta}}(i, x, \bm{\vartheta}, \mu)=(I-\Psi(s_{i}))\bm{\vartheta}(t_{s_{i}})+\Psi(s_{i})\bm{\epsilon}(t_{s_{i}}),
\end{equation}
where, $\Psi(s_{i})=\diag\{\Psi_{1}(s_{i}), \ldots, \Psi_{l}(s_{i})\}$ and $\Psi_{j}(s_{i})\in\mathbb{R}^{n_{j}\times n_{j}}$, $\sum^{l}_{j=1}n_{j}=n_{\bm{\vartheta}}$. If $s_{i}=j$, $\Psi_{j}(s_{i})=I_{n_{j}}$; otherwise, $\Psi_{j}(s_{i})=0$.

Another protocol is Try-Once-Discard (TOD) protocol, which is a quadratic protocol \cite{Donkers2011stability, Nesic2004input}. For the TOD protocol, the node with a minimum index where the norm of the local network-induced error is the largest is allowed to access to the network. The function $H_{\bm{\vartheta}}$ is given by
\begin{equation}
\label{eqn-43}
H_{\bm{\vartheta}}(i, x, \bm{\vartheta}, \mu)=(I-\Psi(\bm{\vartheta}))\bm{\vartheta}(t_{i})+\Psi(\bm{\vartheta})\bm{\epsilon}(t_{s_{i}}),
\end{equation}
where, $\Psi(\bm{\vartheta})=\diag\{\Psi_{1}(\bm{\vartheta}), \ldots, \Psi_{l}(\bm{\vartheta})\}$, and $\Psi_{j}(\bm{\vartheta})=I_{n_{j}}$ if $\min\left\{\arg\max_{1\leq k\leq l}|\bm{\vartheta}_{k}|\right\}=j$; otherwise, $\Psi_{j}(\bm{\vartheta})=0$.

Before verifying Assumptions \ref{asn-4}-\ref{asn-5}, the Lyapunov uniformly globally exponentially stable (UGES) protocol is presented as follows. The Lyapunov UGES protocol was proposed first in \cite[Definition 7]{Nesic2004input} and extended in \cite{Heemels2009networked, Heemels2010networked, Nesic2009unified}.

\begin{definition}
\label{def-2}
The protocol given by $(H_{\bm{\vartheta}}, H_{\mu})$ is \textit{Lyapunov UGES}, if there exist a function $\bar{W}: \mathbb{R}^{n_{\bm{\vartheta}}}\times\mathbb{R}^{l}\times\mathbb{N}\rightarrow\mathbb{R}_{\geq0}$ with $\bar{W}(\cdot, \cdot, c)$ locally Lipschitz for all $c\in\mathbb{N}$, constants $\alpha_{1\bar{W}}, \alpha_{2\bar{W}}>0$ and $\lambda_{1}\in(0, 1)$ such that for all $\bm{\vartheta}\in\mathbb{R}^{n_{\bm{\vartheta}}}$, $\mu\in\mathbb{R}^{l}, c\in\mathbb{N}$ and $x\in\mathbb{R}^{n_{x}}$,
\begin{align}
\label{eqn-44}
\alpha_{1\bar{W}}|(\bm{\vartheta}, \mu)|\leq \bar{W}(\bm{\vartheta}, \mu, c)&\leq\alpha_{2\bar{W}}|(\bm{\vartheta}, \mu)|, \\
\label{eqn-45}
\bar{W}(H_{\bm{\vartheta}}(c, x, \bm{\vartheta}, \mu), H_{\mu}(c, x, \bm{\vartheta}, \mu), c+1)&\leq\lambda_{1}\bar{W}(\bm{\vartheta}, \mu, c).
\end{align}
\end{definition}

Definition \ref{def-2} is an extension of Condition 1 in \cite{Heemels2009networked} and the results in \cite[Subsections IV-C and IV-D]{Nesic2009unified}. Condition 1 in \cite{Heemels2009networked} guarantees the Lyapunov UGES property for the combined protocol of zoom quantizer and network scheduling. In addition, the Lyapunov UGES property is studied for NQCSs in \cite{Nesic2009unified}, where only zoom quantizer and box quantizer are considered. Therefore, Definition \ref{def-2} provides the Lyapunov UGES property for the combined protocol of the general quantizer and network scheduling.

\subsection{Construction of Lyapunov Functions}
\label{subsec-Lyapunov}

Based on the aforementioned Lyapunov UGES protocol, the Lyapunov function satisfying Assumptions \ref{asn-4}-\ref{asn-5} is constructed. To begin with, the following assumption is required.

\begin{assumption}
\label{asn-7}
There exist constants $\lambda_{2}\geq1$, $M_{1}>0$ such that for all $\bm{\vartheta}\in\mathbb{R}^{n_{\bm{\vartheta}}}$, $\mu\in\mathbb{R}^{l}$ and $c\in\mathbb{N}$,
\begin{equation}
\label{eqn-46}
\bar{W}(\bm{\vartheta}, \mu, c+1)\leq\lambda_{2}\bar{W}(\bm{\vartheta}, \mu, c),
\end{equation}
and for almost all $\bm{\vartheta}\in\mathbb{R}^{n_{\bm{\vartheta}}}$, $\mu\in\mathbb{R}^{l}$ and all $c\in\mathbb{N}$,
\begin{equation}
\label{eqn-47}
\max\left\{\left|\frac{\partial \bar{W}(\bm{\vartheta}, \mu, c)}{\partial\bm{\vartheta}}\right|, \left|\frac{\partial \bar{W}(\bm{\vartheta}, \mu, c)}{\partial\mu}\right|\right\}\leq M_{1}.
\end{equation}
\end{assumption}

In Assumption \ref{asn-7}, it follows from \eqref{eqn-46} that $\bar{W}$ is bounded at the transmission jump. In addition, \eqref{eqn-47} equals to the requirement that $\bar{W}$ is globally Lipschitz in $(e, \mu)$ uniformly for $c$, which is valid for the applied protocols in the existing works and this paper. Furthermore, Assumption \ref{asn-7} generalizes the assumptions for the quantization-free case in \cite{Heemels2010networked} and the case of zoom quantization in \cite{Heemels2009networked}. Besides Assumption \ref{asn-7}, assume further that there exist a function $\bar{m}: \mathbb{R}^{n_{x}}\rightarrow\mathbb{R}_{\geq0}$ and constants $M_{e}, M_{\f}\geq0$ such that
\begin{align}
\label{eqn-48}
|g_{e}(x, e)|+|g_{\mu}(x, e, \mu)|&\leq\bar{m}(x)+M_{e}|(e, \mu)|+M_{\f}|e_{\f}|,
\end{align}
which is called the growth condition; see also \cite{Nesic2004input, Heemels2010networked}.

Based on Assumption \ref{asn-7} and \eqref{eqn-48}, the main result of this section is presented as follows.

\begin{theorem}
\label{thm-2}
Consider the system $\mathcal{S}_{2}$. If the following holds:
\begin{enumerate}[i)]
 \item the protocol $(H_{\bm{\vartheta}}, H_{\mu})$ is Lyapunov UGES with a continuous function $\bar{W}$, which is locally Lipschitz in $(\bm{\vartheta}, \mu)$;
 \item Assumption \ref{asn-7} and \eqref{eqn-48} hold for a function $\bar{m}: \mathbb{R}^{n_{x}}\rightarrow\mathbb{R}_{\geq0}$, and constants $\lambda_{2}\geq1$, $M_{1}>0$, $M_{e}, M_{\f}\geq0$;
 \item $|H_{\bm{\vartheta}, j}(i, x, \bm{\vartheta}, \mu)|\leq|\bm{\vartheta}_{j}|$ holds for all $\bm{\vartheta}\in\mathbb{R}^{n_{\bm{\vartheta}}}$, $\mu\in\mathbb{R}^{l}$, $c\in\mathbb{N}$ and each $j\in\{1, \ldots, n_{\bm{\vartheta}}\}$.
\end{enumerate}
then the following function, which is defined below,
\begin{align*}
W(e, \mu, m_{1}, m_{2}, c, b)&:=(1-b)\max\{\bar{W}(e_{\df}, e_{\ct}, 0, \mu, c), \\
&\quad\ \bar{W}(e_{\df}+m^{\df}_{1}, e_{\ct}+m^{\ct}_{1}, 0, \mu+m_{2}, c)\} \\
&\quad\ +b\max\left\{\frac{\lambda_{1}}{\lambda_{2}}\bar{W}(e_{\df}, e_{\ct}, 0, \mu, c), \right.\\
&\quad\ \bar{W}(e_{\df}+m^{\df}_{1}, e_{\ct}+m^{\ct}_{1}, 0, \mu+m_{2}, c)\}
\end{align*}
satisfies Assumptions \ref{asn-4}-\ref{asn-5}. In addition, $\lambda=\lambda_{1}$, $\alpha_{1W}(v)=\alpha_{1\bar{W}}v$, $\alpha_{2W}(v)=\alpha_{2\bar{W}}v$, $\alpha_{3W}(v)=(1+\lambda)M_{1}v$, $\alpha_{4W}(v)=0$, $L_{0}=\alpha^{-1}_{1\bar{W}}M_{1}M_{e}$, $L_{1}=(\lambda_{1}\alpha_{1\bar{W}})^{-1}\lambda_{2}M_{1}M_{e}$, $H_{0}(v)=H_{1}(v)=M_{1}\bar{m}(v)$, and $\sigma_{1W}(v)=\sigma_{2W}(v)=(M_{e}M_{1}/\alpha_{1\bar{W}}+M_{\f})M_{1}v$.
\end{theorem}

\begin{IEEEproof}
According to the local Lipschitz property of $\bar{W}$ and Lebourg's Lipschitz mean value theorem \cite[Theorem 2.3.7]{Clarke1990optimization}, it follows that
\begin{align}
\label{eqn-49}
|(e, \mu)|&\leq|(\bm{\vartheta}, \mu)| \nonumber \\
&\overset{\eqref{eqn-44}}{\leq} [\bar{W}(\bm{\vartheta}, \mu, c)-\bar{W}(e_{\df}, e_{\ct}, 0, \mu, c)]/\alpha_{1\bar{W}} \nonumber \\
&\quad+\bar{W}(e_{\df}, e_{\ct}, 0, \mu, c)/\alpha_{1\bar{W}} \nonumber \\
&\overset{\eqref{eqn-47}}{\leq}\bar{W}(e_{\df}, e_{\ct}, 0, \mu, c)/\alpha_{1\bar{W}}+M_{1}|e_{\f}|/\alpha_{1\bar{W}}.
\end{align}
Define $\mathcal{M}_{1}:=(M_{e}M_{1}/\alpha_{1\bar{W}}+M_{\f})M_{1}$. Based on $W$ defined in Theorem \ref{thm-2}, there are following four cases for Assumptions \ref{asn-4}-\ref{asn-5}.

\textbf{Case 1:} $b=1$ and $\lambda_{1}\bar{W}(e_{\df}, e_{\ct}, 0, \mu, c)\geq\lambda_{2}\bar{W}(e_{\df}+m^{\df}_{1}, e_{\ct}+m^{\ct}_{1}, 0, \mu, c)$. In this case, for the discrete-time instants, we have
\begin{align*}
&W(e, \mu, H_{e}(c, x, e, \mu)-e, H_{\mu}(c, x, e, \mu)-\mu, c+1, 0)\\
&=\frac{\lambda_{1}}{\lambda_{2}}\bar{W}(e_{\df}, e_{\ct}, 0, \mu, c+1)\\
&=\frac{\lambda_{1}}{\lambda_{2}}\left[\bar{W}(e_{\df}, e_{\ct}, 0, \mu, c+1)-\bar{W}(\bm{\vartheta}, \mu, c+1)\right.\\
&\quad \left.+\bar{W}(\bm{\vartheta}, \mu, c+1)\right]-\lambda_{1}\bar{W}(e_{\df}, e_{\ct}, 0, \mu, c) \\
&\quad +\lambda_{1}\bar{W}(e_{\df}, e_{\ct}, 0, \mu, c)\\
&\overset{\eqref{eqn-46}, \eqref{eqn-47}}{\leq}\lambda_{1}\bar{W}(e_{\df}, e_{\ct}, 0, \mu, c)+\lambda_{1}(1+\lambda^{-1}_{2})|e_{\f}|.
\end{align*}
For the continuous-time intervals, we have that
\begin{align*}
\mathcal{I}(1)&=\frac{\lambda_{1}}{\lambda_{2}}\left\langle\frac{\partial\bar{W}(e_{\df}, e_{c}, 0, \mu, c)}{\partial e_{1}}, g_{1}(x, e)\right\rangle\\
&\quad +\frac{\lambda_{1}}{\lambda_{2}}\left\langle\frac{\partial\bar{W}(e_{\df}, e_{c}, 0, \mu, c)}{\partial\mu}, g_{\mu}(x, e, \mu)\right\rangle\\
&\overset{\eqref{eqn-47}, \eqref{eqn-48}}{\leq} \frac{\lambda_{1}}{\lambda_{2}}M_{1}[\bar{m}(x)+M_{e}|(e, \mu)|+M_{\f}|e_{\f}|]\\
&\overset{\eqref{eqn-49}}{\leq}\frac{M_{1}M_{e}}{\alpha_{1\bar{W}}}W(e, \mu, m_{1}, m_{2}, c, 1)+\frac{\lambda_{1}}{\lambda_{2}}M_{1}\bar{m}(x)\\
&\quad +\lambda_{1}\mathcal{M}_{1}|e_{\f}|/\lambda_{2}.
\end{align*}

\textbf{Case 2:} $b=1$ and $\lambda_{1}\bar{W}(e_{\df}, e_{\ct}, 0, \mu, c)\leq\lambda_{2}\bar{W}(e_{\df}+m^{\df}_{1}, e_{\ct}+m^{\ct}_{1}, 0, \mu, c)$. In this case, we yield that
\begin{align*}
& W(e, \mu, H_{e}(c, x, e, \mu)-e, H_{\mu}(c, x, e, \mu)-\mu, c+1, 0)\\
&=\bar{W}(H_{\df}(c, x, e, \mu), H_{\ct}(c, x, e, \mu), 0, H_{\mu}(c, x, \bm{\vartheta}, \mu), c+1)\\
&\quad-\bar{W}(H_{\bm{\vartheta}}(c, x, \bm{\vartheta}, \mu), H_{\mu}(c, x, \bm{\vartheta}, \mu), c+1)\\
&\quad+\bar{W}(H_{\bm{\vartheta}}(c, x, \bm{\vartheta}, \mu), H_{\mu}(c, x, \bm{\vartheta}, \mu), c+1)\\
&\overset{\eqref{eqn-47}}{\leq} M_{1}|H_{\f}(c, x, \bm{\vartheta}, \mu)|\\
&\quad+\bar{W}(H_{\bm{\vartheta}}(c, x, \bm{\vartheta}, \mu), H_{\mu}(c, x, \bm{\vartheta}, \mu), c+1)\\
&\quad-\lambda_{1}\bar{W}(e_{\df}, e_{\ct}, 0, \mu, c)+\lambda_{1}\bar{W}(e_{\df}, e_{\ct}, 0, \mu, c)\\
&\overset{\eqref{eqn-45}, \eqref{eqn-47}}{\leq}\lambda_{1}\bar{W}(e_{\df}, e_{\ct}, 0, \mu, c)+(1+\lambda_{1})M_{1}|e_{\f}|,
\end{align*}
and for the continuous-time intervals, it follows that
\begin{align*}
\mathcal{I}(1)&=\left\langle\frac{\partial\bar{W}(e_{\df}+m^{\df}_{1}, e_{\ct}+m^{\ct}_{1}, 0, \mu, c)}{\partial e_{1}}, g_{1}(x, e)\right\rangle\\
&\quad +\left\langle\frac{\partial\bar{W}(e_{\df}+m^{\df}_{1}, e_{\ct}+m^{\ct}_{1}, 0, \mu, c)}{\partial\mu}, g_{\mu}(x, e, \mu)\right\rangle\\
&\overset{\eqref{eqn-47}, \eqref{eqn-48}}{\leq}M_{1}[\bar{m}(x)+M_{e}|(e, \mu)|+M_{\f}|e_{\f}|]\\
&\overset{\eqref{eqn-49}}{\leq}\frac{\lambda_{2}M_{1}M_{e}}{\lambda_{1}\alpha_{1\bar{W}}}W(e, \mu, m_{1}, m_{2}, c, 1)+M_{1}\bar{m}(x)+\mathcal{M}_{1}|e_{\f}|.
\end{align*}

\textbf{Case 3:} $b=0$ and $\bar{W}(e_{\df}, e_{\ct}, 0, \mu, c)\geq\bar{W}(e_{\df}+m^{\df}_{1}, 0, e_{\ct}+m^{\ct}_{1}, 0, \mu+m_{2}, c)$. At the discrete-time instants, one has that $W(e+m_{1}, \mu+m_{2}, -e-m_{1}, -\mu-m_{2}, c, 1)=\bar{W}(e_{\df}+m^{\df}_{1}, e_{\ct}+m^{\ct}_{1}, 0, \mu+m_{2}, c)$. In the continuous-time intervals,
\begin{align*}
\mathcal{I}(0)&=\left\langle\frac{\partial\bar{W}(e_{\df}, e_{c}, 0, \mu, c)}{\partial e_{1}}, g_{1}(x, e)\right\rangle\\
&\quad +\left\langle\frac{\partial\bar{W}(e_{\df}, e_{c}, 0, \mu, c)}{\partial\mu}, g_{\mu}(x, e, \mu)\right\rangle\\
&\overset{\eqref{eqn-47}, \eqref{eqn-48}}{\leq} M_{1}[\bar{m}(x)+M_{e}|(e, \mu)|+M_{\f}|e_{\f}|]\\
&\overset{\eqref{eqn-49}}{\leq}\frac{M_{1}M_{e}}{\alpha_{1\bar{W}}}W(e, \mu, m_{1}, m_{2}, c, 1)+M_{1}\bar{m}(x)+\mathcal{M}_{1}|e_{\f}|.
\end{align*}

\textbf{Case 4:} $b=0$ and $\bar{W}(e_{\df}, e_{\ct}, 0, \mu, c)\leq\bar{W}(e_{\df}+m^{\df}_{1}, e_{\ct}+m^{\ct}_{1}, 0, \mu+m_{2}, c)$. We get that at the discrete-time instants, $W(e+m_{1}, \mu+m_{2}, -e-m_{1}, -\mu-m_{2}, c, 1)=\bar{W}(0, 0, 0, 0, 0, c)=0$. In the continuous-time intervals,
\begin{align*}
\mathcal{I}(0)&=\left\langle\frac{\partial\bar{W}(e_{\df}+m^{\df}_{1}, e_{\ct}+m^{\ct}_{1}, 0, \mu, c)}{\partial e_{1}}, g_{1}(x, e)\right\rangle\\
&\quad +\left\langle\frac{\partial\bar{W}(e_{\df}+m^{\df}_{1}, e_{\ct}+m^{\ct}_{1}, 0, \mu, c)}{\partial\mu}, g_{\mu}(x, e, \mu)\right\rangle\\
&\overset{\eqref{eqn-43}, \eqref{eqn-46}, \eqref{eqn-47}}{\leq}M_{1}[\bar{m}(x)+M_{e}|(e, \mu)|+M_{\f}|e_{\f}|]\\
&\overset{\eqref{eqn-49}}{\leq}\frac{M_{1}M_{e}}{\alpha_{1\bar{W}}}W(e, \mu, m_{1}, m_{2}, c, 1)+M_{1}\bar{m}(x)+\mathcal{M}_{1}|e_{\f}|.
\end{align*}

Therefore, for $b\in\{0, 1\}$, Assumption \ref{asn-4}-\ref{asn-5} are verified with the parameters given in Theorem \ref{thm-2}. In addition, since \eqref{eqn-44} holds and $W$ is given in Theorem \ref{thm-2}, the bounds on $W$ as in \eqref{eqn-20}-\eqref{eqn-21} are easily obtained. This completes the proof.
\end{IEEEproof}

\begin{remark}
\label{rmk-10}
In Theorem \ref{thm-2}, only the function $W$ can be constructed for the system $\mathcal{S}_{2}$ and the function $V$ can not, which is different from the existing works \cite{Carnevale2007lyapunov, Heemels2010networked, Heemels2009networked}, where both $V$ and $W$ were constructed. The reasons lie in that the objective of this paper is the tracking performance and that $V$ involves in the information of the tracking error, the reference system and the feedback controller. In the previous works \cite{Carnevale2007lyapunov, Heemels2010networked, Heemels2009networked}, the objectives are on stability analysis of NCSs and $V$ is only related to the state of the plant. Since the stability analysis is a basic topic for NCSs and numerous preliminary results have been obtained, $V$ has been constructed in the previous works, whereas $V$ in this paper cannot be developed along the similar techniques as applied in \cite[Section \uppercase\expandafter{\romannumeral5}]{Carnevale2007lyapunov, Heemels2010networked}.
\hfill $\square$
\end{remark}

In Theorem \ref{thm-2}, the Lyapunov function $\bar{W}$ satisfying the items \romannumeral1) and \romannumeral2) exists; see \cite{Nesic2004input, Heemels2009networked, Heemels2010networked, Carnevale2007lyapunov}. The item \romannumeral3) guarantees that the local errors do not increase at each transmission time for all the relevant protocols. Such constraint is valid for all protocols discussed in \cite{Nesic2009unified, Nesic2004input}. We can also double the bound of the local errors due to the quantization error. In the following, two types of the quantizers are introduced as the special cases of the quantizer satisfying Assumption \ref{asn-2} and specific Lyapunov functions are constructed for different cases of time-scheduling protocols and quantizers.

\subsection{Zoom Quantizer}
\label{subsec-zoomquantizer}

A zoom quantizer is defined as
\begin{equation}
\label{eqn-50}
q_{j}(\mu_{j}, z_{j})=\mu_{j}q_{j}\left(\frac{z_{j}}{\mu_{j}}\right),
\end{equation}
where $\mu_{j}>0$, $q_{j}$ in the right-hand side of \eqref{eqn-50} is a uniform quantizer; see \cite{Liu2015dynamic, Liberzon2003hybrid} and Remark \ref{rmk-3} in Section \ref{sec-problemformation}. According to Assumption \ref{asn-2}, we have that for zoom quantizer, $\mathds{C}=\{z=(z_{1}, \ldots, z_{l})\in\mathbb{R}^{n_{z}}| |z_{j}|\leq M_{j}\mu_{j}\}$, $\mathds{D}=\{\epsilon=(\epsilon_{1}, \ldots, \epsilon_{l})\in\mathbb{R}^{n_{z}}| |\epsilon_{j}|\leq\Delta_{j}\mu_{j}\}$ and $\mathds{C}_{0}=\{0\}$. In addition, the quantization parameter is time-invariant in the arrival intervals and updated at the arrival times according to $\mu_{j}(r^{+}_{i})=\Omega_{j}\mu_{j}(r_{i})$ with $\Omega_{j}\in(0, 1)$.

If the zoom quantizer is implemented, then appropriate Lyapunov functions satisfying Assumption \ref{asn-4} are constructed in the following propositions for different time-scheduling protocols.

\begin{proposition}
\label{prop-2}
Let Assumption \ref{asn-3} hold. If the quantizer satisfying Assumption \ref{asn-2} is the zoom quantizer and the protocol in \eqref{eqn-41} is the RR protocol, then Assumption \ref{asn-4} is verified with $W(e, \mu, m_{1}, m_{2}, c, b)=|\mu|+\varpi\sqrt{\sum^{\infty}_{i=c}|\phi_{1}(i, c, e_{1})|^{2}}$, where $\phi_{1}(i, c, e_{1})$ is the solution to $e^{+}_{1}=(h_{\df}(i, x, e_{\df}, \mu), h_{\ct}(i, x, e_{\ct}, \mu))$ at time $i$ starting at time $c$ with the initial condition $e_{1}$, and $\varpi\in(0, (1-\max_{j}\Omega_{j})/(\sqrt{l}\max_{j}\Delta_{j}))$. Moreover, $\lambda=\max\{\sqrt{(l-1)/l}, \varpi\sqrt{l}\max_{j}\Delta_{j}+\max_{j}\Omega_{j}\}$, $\alpha_{1W}(v)=\min\{1, \varpi\}v$, $\alpha_{2W}(v)=(1+\varpi\sqrt{l})v$, $\alpha_{3W}(v)=\alpha_{4W}(v)\equiv0$, $\lambda_{2}=\sqrt{l}$ and $M_{1}=\varpi\sqrt{l}$.
\end{proposition}

\begin{IEEEproof}
For the RR protocol and zoom quantizer case, the Lyapunov function is constructed in several steps. First, the Lyapunov functions for the $e$-subsystem and $\mu$-subsystem are constructed, respectively. Then, such two Lyapunov functions are combined into a Lyapunov function for the whole system.

For the RR protocol, the functions $h_{\df}, h_{\ct}, h_{\f}$ are of the form \eqref{eqn-46}. First, we consider the following system
\begin{align}
\label{eqn-51}
e^{+}_{1}&=\begin{bmatrix}
h_{\df}(c, x, e_{\df}, \mu)\\
h_{\ct}(c, x, e_{\ct}, \mu)
\end{bmatrix}=\begin{bmatrix}
(I-\Psi_{\df}(c))e_{\df}+\Psi_{\df}(c)\epsilon_{\df}\\
(I-\Psi_{\ct}(c))e_{\ct}+\Psi_{\ct}(c)\epsilon_{\ct}
\end{bmatrix}\nonumber\\
&=:H_{1}(c, x, e_{1}, \mu).
\end{align}
Denote by $\phi_{1}(i, c, e_{1})$ the solution to the system \eqref{eqn-51} at time $i$ starting at time $c$ with the initial condition $e_{1}$ and define $W_{1}(e_{1}, c):=\sqrt{\sum^{\infty}_{i=c}|\phi_{1}(i, c, e_{1})|^{2}}$. Based on Proposition 4 in \cite{Nesic2004input} and similar to the proof of Proposition 3 in \cite{Postoyan2014tracking}, we have $|e_{1}|\leq W_{1}(e_{1}, c)\leq\sqrt{l}|e_{1}|$. Define $W_{2}(\mu, c):=|\mu|$ and $W(e, \mu, m_{1}, m_{2}, c, b):=\varpi W_{1}(e_{1}, c)+W_{2}(\mu, c)$, where $\varpi$ is given in Proposition \ref{prop-2}. Thus, we have that \eqref{eqn-20}-\eqref{eqn-21} hold with $\alpha_{1W}(v):=\min\{1, \varpi\}v$ and $\alpha_{2W}(v):=(1+\varpi\sqrt{l})v$, respectively. In addition, $|\bm{\epsilon}|\leq\max_{j}\Delta_{j}|\mu|$ and $|\mu^{+}|\leq\max_{j}\Omega_{j}|\mu|$.

Observe that $\Omega_{j}\in(0, 1)$ and the system \eqref{eqn-51} is dead-beat stable in $l$ steps (see Proposition 4 in \cite{Nesic2004input}), then \eqref{eqn-23} holds with $\alpha_{4W}(v)=0$. In the following, we just need to prove \eqref{eqn-22}. Because of the solution $\phi_{1}$ of the system \eqref{eqn-51} and following from the proof of Proposition 7 in \cite{Nesic2009unified}, we have
\begin{align}
\label{eqn-52}
&W_{1}(H_{1}(c, x, e_{1}, \mu), c+1)\nonumber\\
& \leq\sqrt{\frac{l-1}{l}}W_{1}(e_{1}, c)+\sqrt{l}\max_{j}\Delta_{j}|\mu|.
\end{align}

Combining \eqref{eqn-52} and the fact that $W_{2}(H_{\mu}(c, x, e, \mu))\leq\max_{j}\Omega_{j}|\mu|$ yields that
\begin{align*}
&\varpi W_{1}(H_{1}(c, x, e_{1}, \mu), c+1)+W_{2}(H_{\mu}(c, x, e, \mu), c+1)\\
&\leq\varpi\sqrt{\frac{l-1}{l}}W_{1}(e_{1}, c)+\varpi\sqrt{l}\max_{j}\Delta_{j}W_{2}(\mu)+\max_{j}\Omega_{j}W_{2}(\mu)\\
&\leq\lambda W(e_{1}, \mu, c),
\end{align*}
where $\lambda$ is given in Proposition \ref{prop-2}. Thus, the proof is completed.
\end{IEEEproof}

The following proposition presents the Lyapunov function satisfying Assumption \ref{asn-4} under the TOD protocol and zoom quantizer case. Its proof is a combination of the proofs of Proposition \ref{prop-2} and Proposition 8 in \cite{Nesic2009unified}, and hence omitted here.

\begin{proposition}
\label{prop-3}
Let Assumption \ref{asn-3} hold. If the quantizer satisfying Assumption \ref{asn-2} is the zoom quantizer and the protocol in \eqref{eqn-41} is the TOD protocol, then Assumption \ref{asn-4} is verified with $W(e, \mu, m_{1}, m_{2}, c, b)=\varpi|e_{1}|+|\mu|$, where $\varpi\in(0, (1-\max_{j}\Omega_{j})/(\max_{j}\Delta_{j}))$. Moreover, $\lambda=\max\{\sqrt{(l-1)/l}, \varpi\max_{j}\Delta_{j}+\max_{j}\Omega_{j}\}$, $\alpha_{1W}(v)=\min\{1, \varpi\}v$, $\alpha_{2W}(v)=(1+\varpi)v$, $\alpha_{3W}(v)=\alpha_{4W}(v)\equiv0$, $\lambda_{2}=1$ and $M_{1}=\varpi$.
\end{proposition}

In terms of Theorem \ref{thm-2}, $\alpha_{3W}=0$ is not given \emph{a priori}. However, $\alpha_{3W}\equiv0$ in Propositions \ref{prop-2}-\ref{prop-3}, which implies that the effects of $e_{\f}$ on $\eta$ can be ignored. In this case, the feedforward input can be transmitted to the plant and the reference system directly. To reduce the effects of $\eta$, some addition conditions are needed or the time-scheduling protocol needs to be changed. In the following, we introduce the TOD-tracking protocol \cite{Postoyan2014tracking}, which is a refined TOD protocol with $\Psi_{j}(\bm{\vartheta})$ defined as
\begin{align}
\label{eqn-53}
\Psi_{j}(\bm{\vartheta})=\begin{cases}
I_{n_{j}}, & \text{if } j=\min\{\arg\max_{j}|(e_{\df}, e_{\ct}-e_{\f})_{j}|\}, \\
0, & \text{otherwise}.
\end{cases}
\end{align}
That is, the node to access to the network depends on $(e_{\df}, e_{\ct}-e_{\f})$ instead of $\bm{\vartheta}$. Thus, $W(e, \mu, m_{1}, m_{2}, c, b)=\varpi|(e_{\df}, e_{\ct}-e_{\f})|+|\mu|$ and \eqref{eqn-28} holds with $(\eta, e_{\df}, e_{\ct}-e_{\f}, \mu)$. The TOD-tracking protocol depends on the network setup. For instance, if $u_{\f}$ is generated by the controller \cite{Van2010tracking}, then which node is granted to access the network is associated to $(e_{\df}, e_{\ct}+e_{\f})$. If $y_{\p}$ and $y_{\rf}$ are transmitted via the same nodes, then the node to access to the network is related to $(e_{1}, e_{\f})$. For the TOD-tracking protocol, the following proposition is presented.

\begin{proposition}
\label{prop-4}
Let Assumption \ref{asn-3} hold. If the quantizer satisfying Assumption \ref{asn-2} is the zoom quantizer and the protocol in \eqref{eqn-41} is the TOD-tracking protocol, then Assumption \ref{asn-4} is verified with $W(e, \mu, m_{1}, m_{2}, c, b)=\varpi|(e_{\df}, e_{\ct}-e_{\f})|+|\mu|$, where $\varpi\in(0, (1-\max_{j}\Omega_{j})/(\max_{j}\Delta_{j}))$. Moreover, $\lambda=\max\{\sqrt{(l-1)/l}, \varpi\max_{j}\Delta_{j}+\max_{j}\Omega_{j}\}$, $\alpha_{1W}(v)=\min\{1, \varpi\}v$, $\alpha_{2W}(v)=(1+\varpi)v$, $\alpha_{3W}(v)=\alpha_{4W}(v)\equiv0$, $\lambda_{2}=1$ and $M_{1}=\varpi$.
\end{proposition}

\subsection{Box Quantizer}
\label{subsec-boxquantizer}

Box quantizer, whose quantization regions are rectangle boxes, is another type of dynamical quantizers; see \cite{Liberzon2003stabilization, Nesic2009unified}. A box quantizer is defined by three parameters \cite{Nesic2009unified}: an integer $N_{j}>1$ to define the number of the quantization levels, an estimate $\hat{z}_{j}\in\mathbb{R}^{n_{j}}$ of the variables to be quantized $z_{j}\in\mathbb{R}^{n_{j}}$ and a real number $\mu_{j}\geq0$ to define the size of the quantization regions, where $j\in\{1, \ldots, l\}$. Because $N_{j}$ is a given constant and $\hat{z}_{j}$ depends on $\mu_{j}$, $\mu_{j}$ is called the quantization parameter. Consider the box $\bm{B}(\hat{z}_{j}, \mu_{j})$ and divide it into $N^{n_{j}}_{j}$ equally small sub-boxes numbered from 1 to $N^{n_{j}}_{j}$ in some way. Thus, $q(\mu_{j}, z_{j})$ is the number of the sub-box containing $z_{j}$, and $\hat{z}_{j}$ is updated to be the center of such sub-box.\footnote{If $z_{j}$ is in the intersection of some sub-boxes, then $q(\mu_{j}, z_{j})$ can be any number of these sub-boxes. In addition, we allow $N_{j}$ not to be the same for different nodes, which generalizes the cases in \cite{Nesic2009unified, Liberzon2003stabilization}.} The evolution of $\mu$ is presented as follows; see \cite{Nesic2009unified} for the details.
\begin{align}
\label{eqn-54}
\dot{\mu}(t)&=g_{\mu}(x, e, \mu), \quad t\in(r_{i}, r_{i+1}),\\
\label{eqn-55}
\mu(r^{+}_{i})&=\left(\frac{\mu_{1}(r_{i})}{N_{1}}, \ldots, \frac{\mu_{l}(r_{i})}{N_{l}}\right).
\end{align}

For the RR protocol, $H_{\bm{\vartheta}}$ and $H_{\mu}$ are given by
\begin{align*}
&H_{\bm{\vartheta}}(i, x, \bm{\vartheta}, \mu)=(I-\Psi_{\bm{\vartheta}}(s_{i}))\bm{\vartheta}(t_{s_{i}})+\Psi_{\bm{\vartheta}}(s_{i})J_{\bm{\vartheta}}(s_{i}, x, \bm{\vartheta}, \mu),\\
&H_{\mu}(i, x, \bm{\vartheta}, \mu)=(I-\bar{\Psi}(s_{i}))\mu+N^{-1}\bar{\Psi}(s_{i})\mu,
\end{align*}
where $\Psi_{\bm{\vartheta}}$ and $\bar{\Psi}$ are the diagonal matrices with different dimensions, $N=\diag\{N_{1}, \ldots, N_{l}\}$ and $J_{\bm{\vartheta}}$ depends on the quantization procedure; see \cite{Nesic2009unified}. For the box quantizer and RR protocol case, the following proposition establishes the existence of Lyapunov function satisfying Assumption \ref{asn-4}.

\begin{proposition}
\label{prop-5}
Let Assumption \ref{asn-3} hold. If the RR protocol and the box quantizer are applied and there exists a constant $d>0$ such that $|J_{e}(s_{i}, x, e, \mu)|\leq d|\mu|$,\footnote{The existence of $d$ depends on the expression of $J_{e}(s_{i}, x, e, \mu)$, which has been given in \cite[Section \uppercase\expandafter{\romannumeral3}-C]{Nesic2009unified} (where quantization is considered but transmission delays are not studied).} then Assumption \ref{asn-4} is verified with $W(e, \mu, m_{1}, m_{2}, c, b)=\varpi\sqrt{\sum^{\infty}_{i=c}|\phi_{1}(i, c, e_{1})|^{2}}+\sqrt{\sum^{\infty}_{i=c}|\phi_{\mu}(i, c, \mu)|^{2}}$, where $\phi_{1}(i, c, e_{1})$ and $\phi_{\mu}(i, c, \mu)$ are the solutions to $e^{+}_{1}=(h_{\df}(i, x, e_{\df}, \mu), h_{\ct}(i, x, e_{\ct}, \mu))$ and $\mu^{+}=h_{\mu}(i, x, e, \mu)$ at time $i$ starting at time $c$ with the initial condition $e$ and $\mu$, respectively. Moreover, $\lambda=\max\{\sqrt{(l-1)/l}, \varpi d\sqrt{l}+\bar{\rho}\}$, $\alpha_{1W}(v)=\min\{1, \varpi\}v$,
\begin{align*}
\alpha_{2W}(v)=\left(\varpi\sqrt{l}+\sqrt{\frac{\bar{N}^{2}l}{\bar{N}^{2}-1}}\right)v,  \quad  \bar{\rho}=\sqrt{\frac{\bar{N}^{2}l-\bar{N}^{2}+1}{\bar{N}^{2}l}},
\end{align*}
$\alpha_{4W}(v)=\alpha_{6W}(v)=0$, $\lambda_{2}=\sqrt{l}$, $M_{1}=\varpi\sqrt{l}$, $\bar{N}=\min_{j}N_{j}$ and $\varpi\in(0, (1-\bar{\rho})/(d\sqrt{l}))$.
\end{proposition}

\begin{IEEEproof}
First, we consider the $\mu$-subsystem. Define $W_{2}(\mu, c):=\sqrt{\sum^{\infty}_{i=c}|\phi_{\mu}(i, c, \mu)|^{2}}$, where $\phi_{\mu}(i, c, \mu)$ is the solution to $\mu^{+}=h_{\mu}(i, x, e, \mu)$ at time $i$ starting at time $c$ with the initial condition $\mu$. Let $\bar{N}=\min_{j}N_{j}$ and following the similar line as the proof of Proposition 5 in \cite{Nesic2009unified}, we have
\begin{align}
\label{eqn-56}
|\mu|\leq W_{2}(\mu, c)&\leq\sqrt{\frac{\bar{N}^{2}l}{\bar{N}^{2}-1}}|\mu|,\\
\label{eqn-57}
W_{2}(\mu^{+}, c+1)&\leq\sqrt{\frac{\bar{N}^{2}l-\bar{N}^{2}+1}{\bar{N}^{2}l}}W_{2}(\mu, c).
\end{align}

Next, for the $e_{1}$-subsystem, consider the following system
\begin{align}
\label{eqn-58}
e^{+}_{1}&=\begin{bmatrix}
h_{\df}(c, x, e_{\df}, \mu)\\
h_{\ct}(c, x, e_{\ct}, \mu)
\end{bmatrix}\nonumber\\
&=\begin{bmatrix}
(I-\Psi_{\df}(c))e_{\df}+\Psi_{\df}(c)J_{\df}(s_{i}, x, e_{1}, \mu)\\
(I-\Psi_{\ct}(c))e_{\ct}+\Psi_{\ct}(c)J_{\ct}(s_{i}, x, e_{\ct}, \mu)
\end{bmatrix}\nonumber\\
&=H_{1}(c, x, e_{1}, \mu).
\end{align}
Define $W_{1}(e_{1}, c):=\sqrt{\sum^{\infty}_{i=c}|\phi_{1}(i, c, e_{1})|^{2}}$, where $\phi_{1}(i, c, e_{1})$ is the solutions to \eqref{eqn-58} at time $i$ starting at time $c$ with the initial condition $e_{1}$. Similar to the proof of Proposition \ref{prop-2}, we have $|e_{1}|\leq W_{1}(e_{1}, c)\leq\sqrt{l}|e|$. Based on the proof of Proposition 5 in \cite{Nesic2009unified} and similar to the proof of Proposition \ref{prop-2}, we obtain that
\begin{align}
\label{eqn-59}
&W_{1}(H_{1}(c, x, e_{1}, \mu), c+1) \nonumber\\
&\leq\sqrt{\frac{l-1}{l}}W_{1}(c, e_{1})+d\sqrt{l}W_{2}(\mu, c).
\end{align}

Finally, define $W(e, \mu, m_{1}, m_{2}, c, b):=\varpi W_{1}(e_{1}, c)+W_{2}(\mu, c)$, and \eqref{eqn-20}-\eqref{eqn-21} hold with $\alpha_{1W}(v)=\min\{1, \varpi\}v$ and $\alpha_{2W}(v)=(\varpi\sqrt{l}+\sqrt{\bar{N}^{2}l/(\bar{N}^{2}-1)})v$. Similar to the proof of Proposition \ref{prop-2}, it follows that \eqref{eqn-23} holds. Combining \eqref{eqn-57} and \eqref{eqn-59} yields that
\begin{align*}
&\varpi W_{1}(H_{1}(c, x, e_{1}, \mu), c+1)+W_{2}(H_{\mu}(c, x, e, \mu), c+1)\\
&\leq\varpi\sqrt{\frac{l-1}{l}}W_{1}(e_{1}, c)+\varpi d\sqrt{l}W_{2}(\mu, c) \\
&\quad +\sqrt{\frac{\bar{N}^{2}l-\bar{N}^{2}+1}{\bar{N}^{2}l}}W_{2}(\mu, c)\\
&\leq\lambda W(e_{1}, \mu, c),
\end{align*}
where $\lambda$ is given in Proposition \ref{prop-5}. Thus, \eqref{eqn-22} holds and the proof is completed.
\end{IEEEproof}
For the TOD protocol and the TOD-tracking protocol, the time-scheduling protocol and the update of $\mu$ are given by
\begin{align*}
H_{\bm{\vartheta}}(i, x, \bm{\vartheta}, \mu)&=(I-\Psi(\bm{\vartheta}))\bm{\vartheta}(t_{s_{i}})+\Psi(\bm{\vartheta})J_{\bm{\vartheta}}(s_{i}, x, \bm{\vartheta}, \mu),\\
H_{\mu}(i, x, \bm{\vartheta}, \mu)&=(I-\bar{\Psi}(\bm{\vartheta}))\mu+N^{-1}\bar{\Psi}(\bm{\vartheta})\mu.
\end{align*}
In above cases, the following propositions are presented. The proofs are proceeded along the similar strategy as the proof of Proposition \ref{prop-5}, and hence omitted here.

\begin{proposition}
\label{prop-6}
Let Assumption \ref{asn-3} hold. If the TOD protocol and the box quantizer are applied and there exists a constant $d>0$ such that $|J_{e}(s_{i}, x, e, \mu)|\leq d|\mu|$, then Assumption \ref{asn-4} is verified with $W(e, \mu, m_{1}, m_{2}, c, b)=\varpi|e_{1}|+|\mu|$, where $\varpi>0$. Moreover,  $\lambda=\max\{\sqrt{(l-1)/l}, \varpi d+\tilde{\rho}\}$, $\alpha_{1W}(v)=\min\{1, \varpi\}v$, $\alpha_{2W}(v)=(1+\varpi\sqrt{l})v$, $\alpha_{3W}(v)=\alpha_{4W}(v)=0$, $M_{1}=\varpi$, where $\varpi\in(0, (1-\bar{\rho})/d)$, $\alpha\in(0, 1)$, $\bar{N}=\min_{j}N_{j}$ and
\begin{align}
\label{eqn-60}
\tilde{\rho}=\max\left\{\sqrt{\frac{l-1}{l}}, \sqrt{\frac{\bar{N}^{2}l-\alpha^{2}\bar{N}^{2}+\alpha}{\bar{N}^{2}l}}\right\}.
\end{align}
\end{proposition}

\begin{proposition}
\label{prop-7}
Let Assumption \ref{asn-3} hold. If the TOD-tracking protocol and the box quantizer are applied and there exists a constant $d>0$ such that $|J_{e}(s_{i}, x, e, \mu)|\leq d|\mu|$, then Assumption \ref{asn-4} is verified with $W(e, \mu, m_{1}, m_{2}, c, b)=\varpi|(e_{\eta}, e_{\ct}-e_{\f})|+|\mu|$, where $\varpi>0$. Moreover, $\lambda=\max\{\sqrt{(l-1)/l}, \varpi d+\tilde{\rho}\}$, $\alpha_{1W}(v)=\min\{1, \varpi\}v$, $\alpha_{2W}(v)=(1+\varpi\sqrt{l})v$, $\alpha_{3W}(v)=\alpha_{4W}(v)=0$, $M_{1}=\varpi$, where $\varpi\in(0, (1-\bar{\rho})/d)$, $\alpha\in(0, 1)$, $\bar{N}=\min_{j}N_{j}$ and $\tilde{\rho}$ is given in \eqref{eqn-60}.
\end{proposition}

\section{Illustrative Example}
\label{sec-illustration}

In this section, the developed results are demonstrated with a numerical example. We will illustrate how Assumptions \ref{asn-4}-\ref{asn-6} could be verified and how the verification of these conditions leads to quantitative tradeoff between $h_{\mati}$ and $h_{\mad}$.

Consider a single-link revolute manipulator modeled by Lagrange dynamics \cite{Lewis1998neural, Sciavicco2012modelling}: $M(\bm{q})\ddot{\bm{q}}+C(\bm{q}, \dot{\bm{q}})\dot{\bm{q}}+g(\bm{q})=\bm{\tau}$. Pick $M(\bm{q})=1$, $C(\bm{q}, \dot{\bm{q}})=0$, $g(\bm{q})=m\cos\bm{q}$, $\bm{\tau}=au$ and define $\bm{q}_{1}=\bm{q}$, $\bm{q}_{2}=\dot{\bm{q}}$, we have that
\begin{align}
\label{eqn-61}
\dot{\bm{q}}_{1}=\bm{q}_{2}, \quad \dot{\bm{q}}_{2}=-m\cos\bm{q}_{1}+au,
\end{align}
where $\bm{q}_{1}$ is the generalized configuration coordinate, $\bm{q}_{2}$ is the generalized velocity, $\bm{\tau}$ is the generalized force acting on the system and $m, a>0$ are certain constants. Both $\bm{q}_{1}$ and $\bm{q}_{2}$ are measurable. The reference system is given by
\begin{align}
\label{eqn-62}
\dot{\bm{q}}_{\rf1}=\bm{q}_{\rf2}, \quad \dot{\bm{q}}_{\rf2}=-m\cos\bm{q}_{\rf1}+au_{\f},
\end{align}
where $\bm{q}_{\rf1}, \bm{q}_{\rf2}$ are measurable and $u_{\f}=2\cos(5t)$.

Assume that if there is no communication network, then the tracking error is asymptotically stable. The designed controller is $u=u_{\f}+u_{\ct}$, where $u_{\ct}=-a^{-1}[m\sin(\bm{q}_{\df1}/2)+\bm{q}_{\df1}+\bm{q}_{\df2}]$, where $\bm{q}_{\df1}:=\bm{q}_{1}-\bm{q}_{\rf1}$ and $\bm{q}_{\df2}:=\bm{q}_{2}-\bm{q}_{\rf2}$. In this paper, we consider the case that the communication between the controller and the plant is via a quantizer and a communication network. The applied quantizer is zoom quantizer \eqref{eqn-50} with $\max_{j}\Delta_{j}=0.8$ and $\max_{j}\Omega_{j}=0.6$. The controller is applied using ZOH devices and the network is assumed to have $l=3$ nodes for $\bm{q}_{\df1}$, $\bm{q}_{\df2}$ and $u$, respectively. In this case, the applied feedback controller is emulated and given by $u_{\ct}=-a^{-1}[m\sin(\hat{\bm{q}}_{\df1}/2)+\hat{\bm{q}}_{\df1}+\hat{\bm{q}}_{\df2}]$. In addition, $u_{\f}$ is assumed to be transmitted to the reference system directly for the TOD-tracking protocol and $\hat{\bm{q}}_{\df1}, \hat{\bm{q}}_{\df2}$ are implemented in ZOH fashion in the update intervals, which implies that the feedback controller knows $\bm{q}_{\rf1}, \bm{q}_{\rf2}$ but does not depend on $\bm{q}_{\rf1}, \bm{q}_{\rf2}$. To simplify the following simulation, the transmission intervals and the transmission delays are constants, that is, $h_{i}\equiv h_{\mati}$ and $\tau_{i}\equiv h_{\mad}$ for all $i\in\mathbb{N}$.

\begin{figure}[!t]
\begin{center}
\begin{picture}(90,120)
\put(-55,-5){\resizebox{70mm}{45mm}{\includegraphics[width=2.5in]{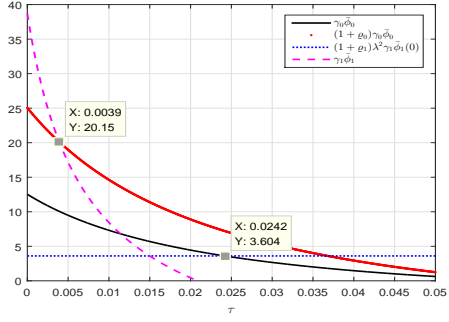}}}
\end{picture}
\end{center}
\caption{The functions $\bar{\phi}_{b}$, $b\in\{0, 1\}$ with $\bar{\phi}_{0}(0)=\bar{\phi}_{1}(0)=\sqrt{3}$ for RR protocol.}
\label{fig-2}
\end{figure}

\begin{figure}[!t]
\begin{center}
\begin{picture}(90,120)
\put(-55,-5){\resizebox{70mm}{45mm}{\includegraphics[width=2.5in]{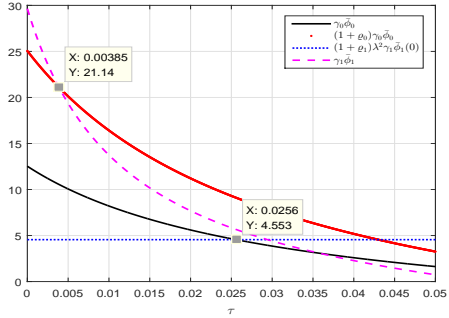}}}
\end{picture}
\end{center}
\caption{The functions $\bar{\phi}_{b}$, $b\in\{0, 1\}$ with $\bar{\phi}_{0}(0)=\sqrt{3}$, $\bar{\phi}_{1}(0)=\sqrt{3}+1$ for TOD and TOD-tracking protocols.}
\label{fig-3}
\end{figure}

In the following, we first verify Assumptions \ref{asn-4}-\ref{asn-6}. Based on \eqref{eqn-61}-\eqref{eqn-62}, we obtain that $F_{\eta}=
(\eta_{2}, -m[\cos(\eta_{1}+e_{\df1}+\bm{q}_{\rf1})-\cos(\bm{q}_{\rf1})+\sin((\eta_{1}+e_{\df1})/2)]-(\eta_{1}+e_{\df1})-(\eta_{2}+e_{\df2})+au_{\f}+ae_{\f})$, $F_{\rf}=(\bm{q}_{\rf2}, -m\cos\bm{q}_{\rf1}+au_{\f}+ae_{\f})$, $G_{1}=(-F_{\eta}, 0)$, $G_{\rf}=-F_{\rf}$ and $G_{\f}=-\dot{u}_{\f}$. Choose the Lyapunov function $W$ in Proposition \ref{prop-2} for the RR protocol, $W(e, \mu, m_{1}, m_{2}, c, b):=\varpi|e|+|\mu|$ for the TOD protocol and $W(e, \mu, m_{1}, m_{2}, c, b):=\varpi|(e_{\df}, e_{\ct}-e_{\f})|+|\mu|$ for the TOD-tracking protocol. Thus, Assumption \ref{asn-4} holds easily based on Propositions \ref{prop-2}-\ref{prop-4}.

On the other hand, it holds that $|g_{1}|\leq|\eta_{2}|+|\eta_{1}+\eta_{2}|+E_{1}|e_{1}|+a|e_{\f}|$, where $E_{1}=m+\sqrt{3}\max\{1, a\}$. It is known from Propositions \ref{prop-2}-\ref{prop-4} that $\alpha_{1W}(v)=\min\{1, \varpi\}v$ for all $v\geq0$ and $\max\{|\partial\bar{W}(\bm{\vartheta}, \mu, c)/\partial\bm{\vartheta}|, |\partial \bar{W}(\bm{\vartheta}, \mu, c)/\partial\mu|\}\leq M_{1}$ for almost all $e, m_{1}\in\mathbb{R}^{n_{e}}, \mu, m_{2}\in\mathbb{R}^{l}, c\in\mathbb{N}$ and $b\in\{0, 1\}$, where $M_{1}=\sqrt{l}$ for the RR protocol and $M_{1}=1$ for the TOD and TOD-tracking protocols. Thus, $|\langle\partial W(e, \mu, m_{1}, m_{2}, c, b)/\partial e, g_{e}(x, e)\rangle|\leq\varpi M_{1}[|\eta_{1}|+|\eta_{1}+\eta_{2}|+E_{1}|e_{1}|+a|e_{\f}|]$. As a result, Assumption \ref{asn-5} holds with $L_{0}=\varpi M_{1}E_{1}/\alpha_{1W}$, $L_{1}=\varpi M_{1}E_{1}\lambda_{2}/(\lambda_{1}\alpha_{1W})$, $H_{0}(x)=H_{1}(x)=\varpi M_{1}(|\eta_{1}|+|\eta_{1}+\eta_{2}|)$ and $\sigma_{bW}(v)=\varpi aM_{1}|v|$.

\begin{figure}[!t]
\begin{center}
\begin{picture}(90,120)
\put(-55,-5){\resizebox{70mm}{45mm}{\includegraphics[width=2.5in]{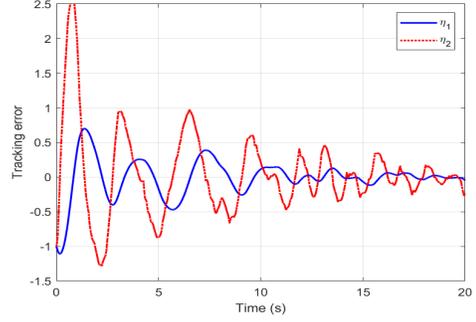}}}
\end{picture}
\end{center}
\caption{Tracking errors for $h_{\mati}=0.0242$, $h_{\mad}=0.00390$ and RR protocol.}
\label{fig-4}
\end{figure}

\begin{figure}[!t]
\begin{center}
\begin{picture}(90,120)
\put(-55,-5){\resizebox{70mm}{45mm}{\includegraphics[width=2.5in]{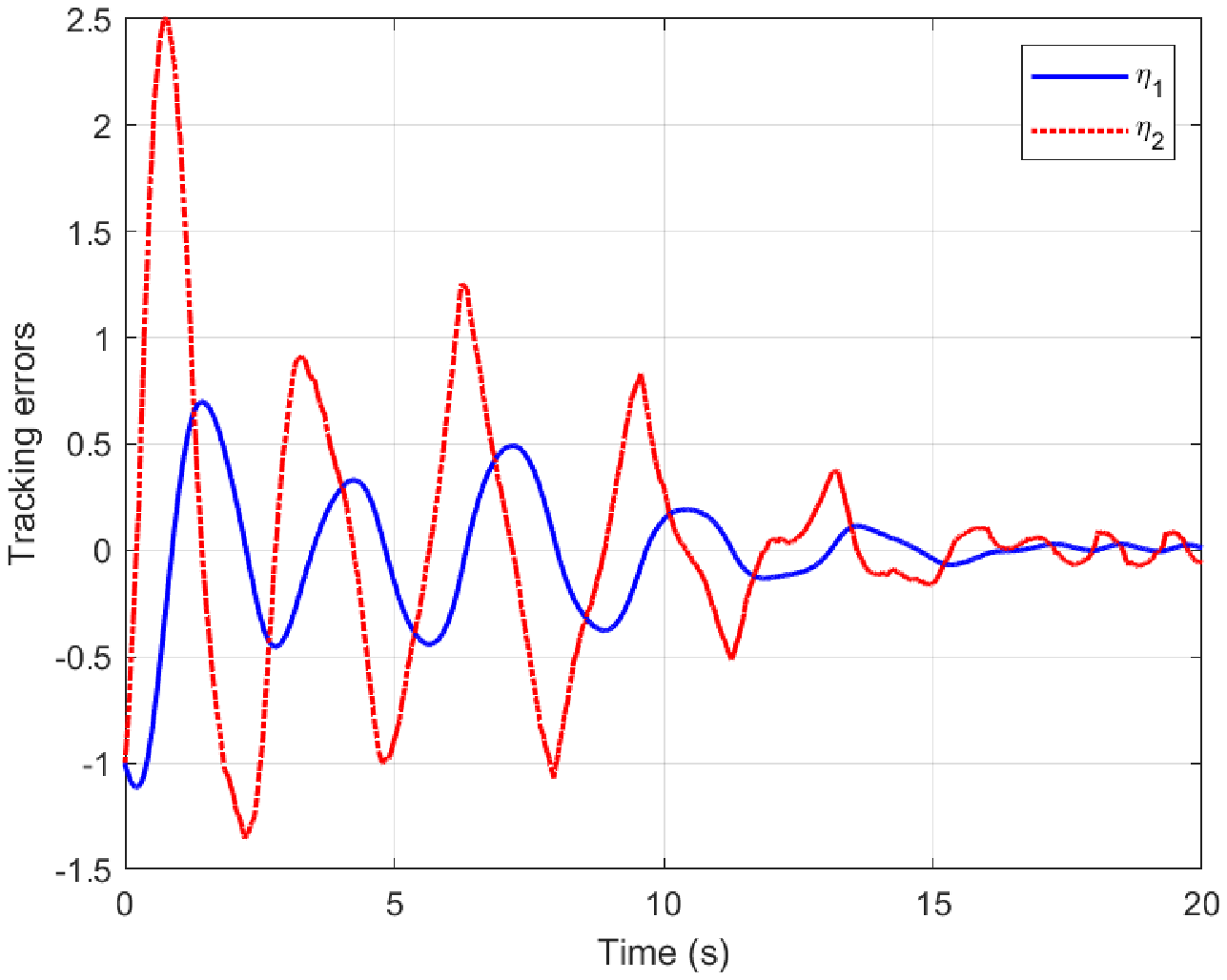}}}
\end{picture}
\end{center}
\caption{Tracking errors for $h_{\mati}=0.0256$, $h_{\mad}=0.00385$ and TOD protocol.}
\label{fig-5}
\end{figure}

To verify Assumption \ref{asn-6}, define $V(\eta):=\phi_{1}\eta^{2}_{1}+\phi_{2}\eta_{1}\eta_{2}+\phi_{3}\eta^{2}_{2}$, where $\phi_{1}, \phi_{2}, \phi_{3}$ are chosen to make $V$ satisfy \eqref{eqn-25}. Assume that there exists a time-varying parameter $\hat{m}\in[-m, m]$ such that $m[\cos(\eta_{1}+e_{\df1}+\bm{q}_{\rf1})-\cos(\bm{q}_{\rf1})+\sin((\eta_{1}+e_{\df1})/2)]=\hat{m}e_{\df1}$. Thus, using twice the fact that $2xy\leq cx^{2}+y^{2}/c$ for all $x, y\geq0$ and $c>0$, we get that $\langle\nabla V(\eta), F_{\eta}(x, e)\rangle\leq-\phi_{2}\eta^{2}_{1}-(2\phi_{3}-\phi_{2})\eta^{2}_{2}+(2\phi_{1}-2\phi_{3}-\phi_{2})\eta_{1}\eta_{2}+(\varrho^{-1}_{0}+\varrho^{-1}_{1})(\phi_{2}\eta_{1}+2\phi_{3}\eta_{2})^{2}
+\varrho_{0}E^{2}_{2}|e_{1}|^{2}+\varrho_{1}a^{2}|e_{\f}|^{2}$, where $\varrho_{0}, \varrho_{1}>0$ are defined in \eqref{eqn-27} and $E_{2}=\sqrt{3}\max\{1+m, a\}$. Therefore, if $\phi_{1}, \phi_{2}, \phi_{3}$ are chosen such that \eqref{eqn-25} holds and
\begin{align}
\label{eqn-63}
&-\rho_{b}(|\eta|)-H^{2}_{b}(x)\geq-\phi_{2}\eta^{2}_{1}+(2\phi_{1}-2\phi_{3}-\phi_{2})\eta_{1}\eta_{2}\nonumber\\
&\quad -(2\phi_{3}-\phi_{2})\eta^{2}_{2}+(\varrho^{-1}_{0}+\varrho^{-1}_{1})(\phi_{2}\eta_{1}+2\phi_{3}\eta_{2})^{2},
\end{align}
then Assumption \ref{asn-6} is verified with $\theta_{b}(v)=\pi v^{2}$, $\gamma_{0}=\sqrt{\pi+\varrho_{0}E^{2}_{2}}$, $\gamma_{1}=\sqrt{\pi+\varrho_{1}\lambda^{2}_{2}E^{2}_{2}/\lambda^{2}_{1}}$, $\sigma_{bV}(v)=\varrho_{1}a^{2}|v|^{2}$ and $\pi>0$ is arbitrarily small.

Observe that $L_{0}$ and $L_{1}$ depend on the magnitude of $m$ and $a$; $\gamma_{0}$ and $\gamma_{1}$ depend on the choice of $\varrho_{0}, \varrho_{1}$ and $\varepsilon$. Thus, the tradeoff between $h_{\mati}$ and $h_{\mad}$ is related to $m, a, \varrho_{0}$ and $\varrho_{1}$. Pick $m=4.905$, $a=2$, $\varpi=\pi=0.005$, $\varrho_{0}=1$ and $\varrho_{1}=\varrho_{0}\lambda_{2}/\lambda_{1}$. The choices of $\phi_{1}, \phi_{2}$ and $\phi_{3}$ are to satisfy \eqref{eqn-25} and \eqref{eqn-63}. Thus, $L_{0}=17.7150$, $L_{1}=37.5792$, $\gamma_{0}=7.2325$, $\gamma_{1}=22.3450$ for the RR protocol and $L_{0}=10.2278$, $L_{1}=21.6964$, $\gamma_{0}=7.2325$, $\gamma_{1}=22.3450$ for the TOD and the TOD-tracking protocols.

\begin{figure}[!t]
\begin{center}
\begin{picture}(90,120)
\put(-55,-5){\resizebox{70mm}{45mm}{\includegraphics[width=2.5in]{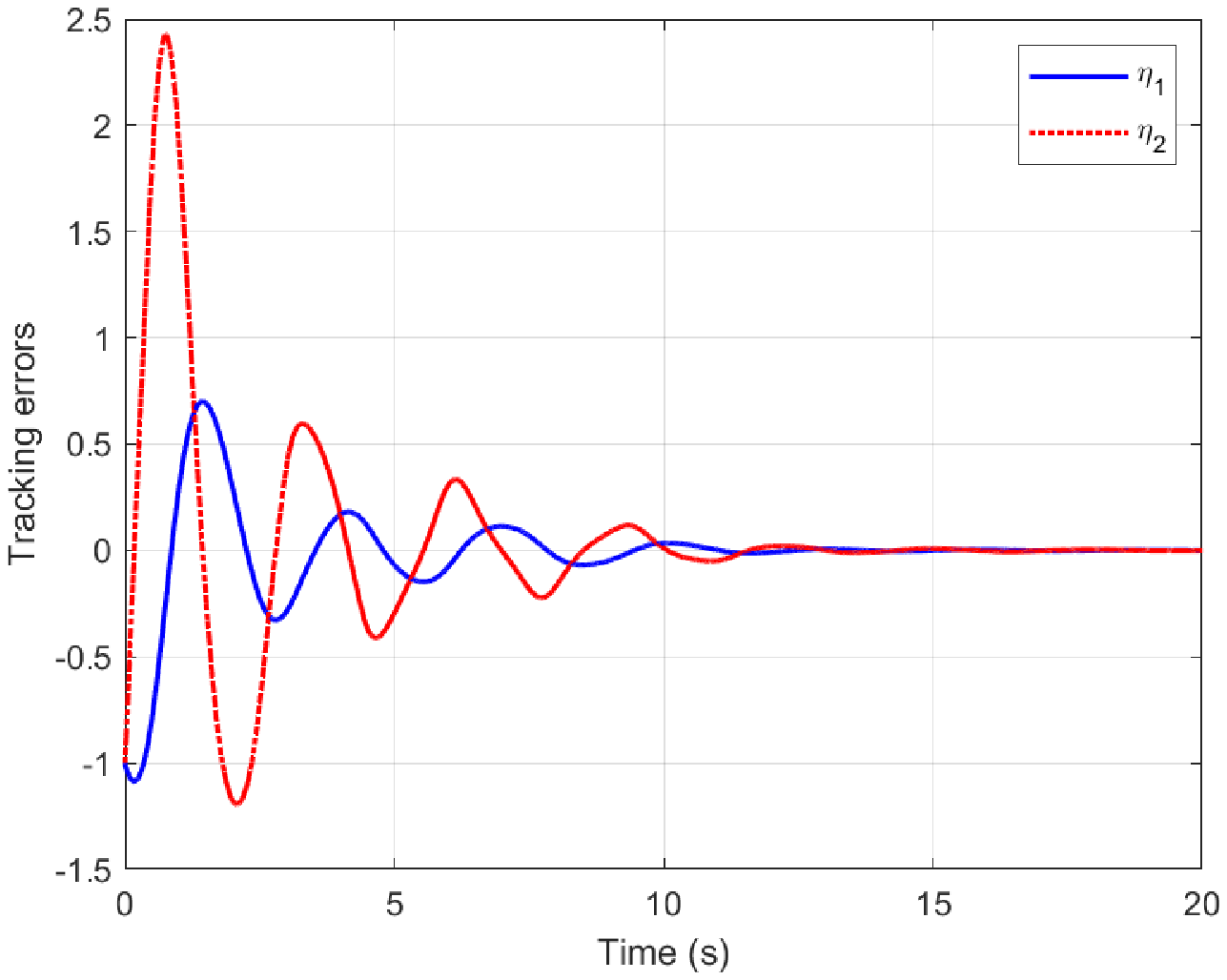}}}
\end{picture}
\end{center}
\caption{Tracking errors for $h_{\mati}=0.0256$, $h_{\mad}=0.00385$ and TOD-tracking protocol.}
\label{fig-6}
\end{figure}

Set the initial $\bar{\phi}_{0}(0)=\bar{\phi}_{1}(0)=\sqrt{3}$ for the RR protocol, and $\bar{\phi}_{0}(0)=\sqrt{3}, \bar{\phi}_{1}(0)=\sqrt{3}+1$ for the TOD and TOD-tracking protocols, then $h_{\mati}$ and $h_{\mad}$ are obtained via \eqref{eqn-28-1}-\eqref{eqn-28-2}; see Figs. \ref{fig-2}-\ref{fig-3}. Follows from Figs. \ref{fig-2}-\ref{fig-3}, $h_{\mati}=0.0242$ and $h_{\mad}=0.00390$ for the RR protocol; $h_{\mati}=0.0256$ and $h_{\mad}=0.00385$ for the TOD and TOD-tracking protocols. In addition, the magnitudes of $h_{\mati}$ and $h_{\mad}$ are different from different initial conditions. For instance, $h_{\mati}=0.0255$ and $h_{\mad}=0.00425$ for the RR protocol with $\bar{\phi}_{0}(0)=\bar{\phi}_{1}(0)=\sqrt{2}$; $h_{\mati}=0.0242$ and $h_{\mad}=0.0069$ for the TOD protocol with $\bar{\phi}_{0}(0)=\sqrt{2}$ and $\bar{\phi}_{1}(0)=\sqrt{2}+1$. $h_{\mati}=0.02375$ and $h_{\mad}=0.00365$ for the RR protocol with $\bar{\phi}_{0}(0)=\bar{\phi}_{1}(0)=2$; $h_{\mati}=0.02615$ and $h_{\mad}=0.0024$ for the TOD protocol with $\bar{\phi}_{0}(0)=2$ and $\bar{\phi}_{1}(0)=3$. Therefore, different tradeoffs between $h_{\mati}$ and $h_{\mad}$ can be obtained for different initial conditions. Furthermore, from \eqref{eqn-28}, $h_{\mati}$ and $h_{\mad}$ depend on $\varrho_{0}$ and $\varrho_{1}$. For instance, changing the value of $\varrho_{0}$ leads to the different magnitudes of $h_{\mati}$ and $h_{\mad}$. The maximal value of $\varrho_{0}$ is 2.090. If $\varrho_{0}>2.090$, then there is no intersection between $\gamma_{1}\bar{\phi}_{1}$ and $(1+\varrho_{0})\gamma_{0}\bar{\phi}_{0}$.

Based on $h_{\mati}$ and $h_{\mad}$ obtained from Figs. \ref{fig-2}-\ref{fig-3}, the tracking errors are illustrated in Figs. \ref{fig-4}-\ref{fig-6} for different cases. Figs. \ref{fig-4}-\ref{fig-5} are for the RR and TOD protocol cases, respectively. The tracking errors are convergent and bounded in Figs. \ref{fig-4}-\ref{fig-5} due to the effects of the network-induced error $e_{\f}$. Since $u_{\f}$ is transmitted to the reference system and the plant directly in the TOD-tracking protocol, $e_{\f}=0$ and the tracking error converge to zero in Fig. \ref{fig-6}.

\section{Conclusions}
\label{sec-conclusion}

In this paper, the tracking control of nonlinear networked and quantized control systems was analyzed based on a Lyapunov approach. To deal with the network-induced issues, a new hybrid model was developed. Sufficient conditions were established to guarantee the convergence of the tracking error. In addition, the Lyapunov function satisfying the obtained conditions was constructed. For the specific time-scheduling protocols and quantizers, we discussed how to reduce the effects of the network-induced errors on the convergence of the tracking error. Finally, the developed results were demonstrated by a numerical example.

In the future, our work will focus on tracking control of networked and quantized control systems with external disturbances. Since the disturbances may lead the system state to escape from the quantization regions, the general quantization mechanism in \cite{Liberzon2003hybrid, Liberzon2003stabilization} needs to be considered.

\section*{Appendix}
\label{sec-appendix}

The detailed expressions of the functions in $\mathcal{S}_{1}$ are presented as follows. As the preceding definitions in Section \ref{sec-problemformation}, we have $x_{\p}=\eta+x_{\rf}$. Moreover, based on \eqref{eqn-2}-\eqref{eqn-5}, \eqref{eqn-10}, we have that for $t\in[t_{s_{i}}, t_{s_{i+1}}]\setminus\{r_{i}\}$, $i\in\mathbb{N}$,
\begin{align*}
\dot{\eta}&=f_{\p}(x_{\p}, \hat{u}_{\ct}+\hat{u}_{\f})-f_{\p}(t, x_{\rf}, \hat{u}_{\f})\\
&=f_{\p}(x_{\p}, u_{\ct}+e_{\ct}+u_{\f}+e_{\f})-f_{\p}(t, x_{\rf}, u_{\f}+e_{\f})\\
&=f_{\p}(\eta+x_{\rf}, g_{\ct}(x_{\ct})+e_{\ct}+u_{\f}+e_{\f})-f_{\p}(t, x_{\rf}, u_{\f}+e_{\f})\\
&=:F_{\eta}(\eta, x_{\ct}, x_{\rf}, e_{1}, e_{\f}), \\
\dot{x}_{\ct}&=f_{\ct}(x_{\ct}, \hat{y}_{\df})=f_{\ct}(x_{\ct}, y_{\df}+e_{\df})\\
&=f_{\ct}(x_{\ct}, g_{\p}(\eta+x_{\rf})-g_{\rf}(x_{\rf})+e_{\df})\\
&=:F_{\ct}(\eta, x_{\ct}, x_{\rf}, e_{1}, e_{\f}),\\
\dot{x}_{\rf}&=f_{\p}(x_{\rf}, \hat{u}_{\f})=f_{\p}(x_{\rf}, u_{\f}+e_{\f})=:F_{\rf}(\eta, x_{\ct}, x_{\rf}, e_{1}, e_{\f}),\\
\dot{\mu}&=g_{\mu}(x_{\p}, x_{\ct}, x_{\rf}, \bm{\vartheta}, \mu)\\
&=g_{\mu}(\eta+x_{\rf}, x_{\ct}, x_{\rf}, e_{\df}, e_{\ct}, e_{\f}, \mu)\\
&=:G_{\mu}(\eta, x_{\ct}, x_{\rf}, e_{1}, e_{\f}, \mu),\\
\dot{e}_{1}&=\begin{bmatrix}
\dot{e}_{\df}\\ \dot{e}_{\ct}
\end{bmatrix}=\begin{bmatrix}
\dot{e}_{\p}-\dot{e}_{\rf} \\ \dot{e}_{\ct}
\end{bmatrix}\\
&=\begin{bmatrix}
\langle\nabla g_{\p}(x_{\rf}), f_{\p}(x_{\rf}, u_{\f}+e_{\f})\rangle \\
-\langle\nabla g_{\p}(x_{\p}), f_{\p}(\eta+x_{\rf}, g_{\ct}(x_{\ct})+e_{\ct}+u_{\f}+e_{\f})\rangle\\
\langle\nabla g_{\ct}(x_{\ct}), f_{\ct}(\eta, x_{\ct}, x_{\rf}, e_{\eta})\rangle
\end{bmatrix}\\
&=:G_{1}(\eta, x_{\ct}, x_{\rf}, e_{1}, e_{\f}) , \\
\dot{e}_{\rf}&=-\langle\nabla g_{\rf}(x_{\rf}), f_{\rf}(x_{\rf}, u_{\f}+e_{\f})\rangle=:G_{\rf}(\eta, x_{\ct}, x_{\rf}, e_{1}, e_{\f}) ,\\
\dot{e}_{\f}&=-\dot{u}_{\f}=-\dfrac{\partial u_{\f}}{\partial x_{\p}}(F_{\eta}+F_{\rf})-\dfrac{\partial u_{\f}}{\partial x_{\rf}}F_{\rf}-\dfrac{\partial u_{\f}}{\partial x_{\ct}}F_{\ct} \\
&=:G_{\f}(\eta, x_{\ct}, x_{\rf}, e_{1}, e_{\f}).
\end{align*}
At the arrival time $r_{i}$, $i\in\mathbb{N}$,
\begin{align*}
h_{\mu}(i, \mu, \bm{\epsilon})&=h_{\mu}(i, \mu, \epsilon_{\df}, \epsilon_{\ct}, \epsilon_{\f})\\
&=h_{\mu}(i, \mu, q(\mu, g_{\p}(x_{\p})-g_{\rf}(x_{\rf}))-g_{\p}(x_{\p})+g_{\rf}(x_{\rf}), \\
&\quad q(\mu, g_{\ct}(x_{\ct}))-g_{\ct}(x_{\ct}), q(\mu, u_{\f})-u_{\f})\\
&=:H_{\mu}(i, \eta, x_{\ct}, x_{\rf}, e_{1}, e_{\f}, \mu),\\
h_{\df}(i, y_{\df}, \bm{\vartheta}, \mu)&=\epsilon_{\df}+\bm{h}_{\df}(i, \bm{\vartheta}) \\
&=q(\mu, g_{\p}(x_{\p})-g_{\rf}(x_{\rf}))-g_{\p}(x_{\p})+g_{\rf}(x_{\rf}) \\
&\quad +\bm{h}_{\df}(i, e_{\df}, e_{\ct}, e_{\f})\\
&=:H_{\df}(i, \eta, x_{\rf}, x_{\ct}, e_{1}, e_{\f}, \mu),\\
h_{\ct}(i, x_{\ct}, \bm{\vartheta}, \mu)&=\epsilon_{\ct}+\bm{h}_{\ct}(i, \bm{\vartheta})\\
&=q(\mu, g_{\ct}(x_{\ct}))-g_{\ct}(x_{\ct})+\bm{h}_{\ct}(i, e_{\df}, e_{\ct}, e_{\f})\\
&=:H_{\ct}(i, \eta, x_{\rf}, x_{\ct}, e_{1}, e_{\f}, \mu),\\
h_{\f}(i, x_{\f}, \bm{\vartheta}, \mu)&=\epsilon_{\f}+\bm{h}_{\f}(i, \bm{\vartheta})\\
&=q(\mu, u_{\f})-u_{\f}+\bm{h}_{\f}(i, e_{\df}, e_{\ct}, e_{\f})\\
&=:H_{\f}(i, \eta, x_{\rf}, x_{\ct}, e_{1}, e_{\f}, \mu).
\end{align*}

\bibliographystyle{IEEEtran}

\begin{IEEEbiography}{Wei Ren}
received his B.Sci. degree from Hubei University, China, and his Ph.D. degree from the University of Science and Technology of China, China, in 2011 and 2018. He was a joint-Ph.D. under the supervision of Professor Dragan Ne{\v{s}}i{\'c} in the University of Melbourne, Victoria, Australia. Currently, he is a post-doctor at KTH Royal Institute of Technology, Stockholm, Sweden. His research interests include networked control systems, nonlinear systems, symbolic abstraction, multi-agent systems and hybrid systems.
\end{IEEEbiography}

\begin{IEEEbiography}{Junlin Xiong}
received his B.Eng. and M.Sci. degrees from Northeastern University, China, and his Ph.D. degree from the University of Hong Kong, China, in 2000, 2003 and 2007, respectively. From November 2007 to February 2010, he was a research associate at the University of New South Wales at the Australian Defense Force Academy, Australia. In March 2010, he joined the University of Science and Technology of China where he is currently a Professor in the Department of Automation. His current research interests are in the fields of Markovian jump systems, networked control systems and negative imaginary systems.
\end{IEEEbiography}

\end{document}